\documentclass[aps,prb,showpacs,twocolumn,10]{revtex4-1}
\pdfoutput=1
\pdfpageattr {/Group << /S /Transparency /I true /CS /DeviceRGB>>}
\usepackage{graphicx}
\usepackage{amsmath}
\usepackage{bm}

\usepackage{hyperref}
\hypersetup{%
pdftitle={Topological insulators are tunable waveguides for hyperbolic polaritons},%
pdfauthor={J.-S. Wu et al.},%
pdfpagemode={UseNone},%
pdfstartview={FitH},%
breaklinks=true,%
citecolor=blue,%
colorlinks=true,%
linkcolor=blue,%
urlcolor=blue}

\def\BiSe*{Bi$_2$Se$_3$}
\def\GHG*{G/hBN/G}
\newcommand{\unit}[1]{\,\mathrm{#1}} 

\usepackage{CJKutf8}	

\begin{document}

\begin{CJK*}{UTF8}{bkai}	
\title{Topological insulators are tunable waveguides for hyperbolic polaritons}

\author{Jhih-Sheng Wu (\CJKchar{"54}{"33}\CJKchar{"81}{"F4}\CJKchar{"76}{"DB}),
D. N. Basov,
and M. M. Fogler}
\affiliation{University of California San Diego, 9500 Gilman Drive, La Jolla, California 92093, USA}


\date{\today}

\begin{abstract}

Layered topological insulators,
for example, \BiSe* are optically hyperbolic materials in a range of THz frequencies.
Such materials possess deeply subdiffractional, highly directional
collective modes:
hyperbolic phonon-polaritons.
In thin crystals
the dispersion of such modes is split into discrete subbands
and is strongly influenced by electron surface states.
If the surface states are doped,
then hybrid collective modes result from coupling
of the phonon-polaritons with surface plasmons.
The strength of the hybridization can be controlled by an external gate that
varies the chemical potential of the surface states.
Momentum-dependence of the plasmon-phonon coupling
leads to a polaritonic analog of the Goos-H\"anchen effect.
Directionality of the polaritonic rays
and their tunable Goos-H\"anchen shift are observable via THz nanoimaging.

\end{abstract}
\maketitle
\end{CJK*} 

\section{Introduction}
\label{sec:Introduction}

\begin{figure}[t]
\includegraphics{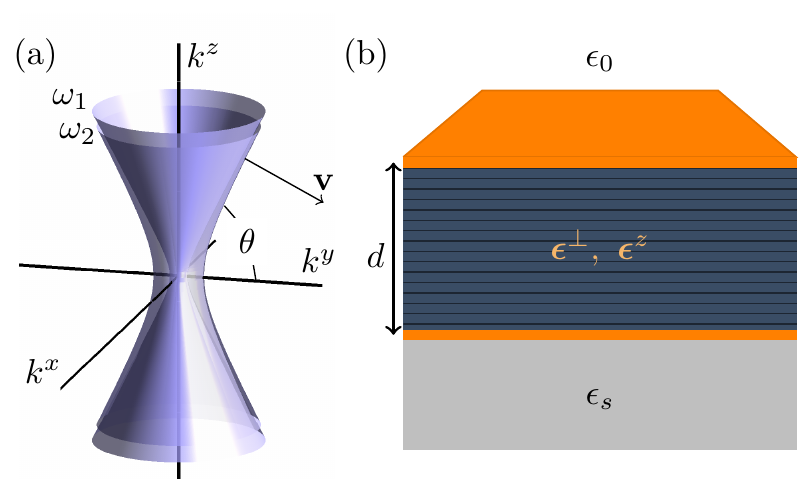}
\caption{(Color online) (a) Hyperboloidal isofrequency surfaces of HP$^2$s for two frequencies $\omega_1$ and $\omega_2$ ($\omega_2 > \omega_1$).
The asymptote angle $\theta$ with respect to the $k^x$--$k^y$ plane is shown; the group velocity $\mathbf{v}$ makes the same angle with respect to the $k^z$-axis.
(b) Model geometry:
a TI slab of thickness $d$ sandwiched between a substrate of permittivity $\epsilon_s$ and a superstrate of permittivity $\epsilon_0$.
The two thin (orange) layers represent the top and the bottom surfaces states.
}
\label{fig:model_b}
\end{figure}

Bismuth-based topological insulators (TIs) have attracted much interest for their unusual electron surface states (SSs),
which behave as massless Dirac fermions.~\cite{Hasan2010, Qi2011tia}
However, bulk optical response of these compounds~\cite{Richter1977, LaForge2010, Cheng2011, Akrap2012, DiPietro2012ocb, DiPietro2013, Wu2013sct, Post2013tdb, Chapler2014ief, Reijnders2014, Autore2015omb, Autore2015ppi, Post2015src} is also remarkable. 
The quintuple-layered structure of these materials causes a strong anisotropy of their phonon modes.
The ${E}_u$ phonons that involve atomic displacements in the plane parallel to the basal plane (henceforth, $x$--$y$ or $\bot$--plane) have lower frequencies than ${A}_{2u}$, the $c$-axis (henceforth, $z$-axis) vibrations.\cite{Cheng2011}
For \BiSe*, the dominant $\bot$- and $z$-axis phonon frequencies, 
\begin{equation}
\begin{alignedat}{2}
\omega_{1,to}^{\bot} &= 64\unit{cm}^{-1} & &= 1.9\unit{THz}\,,
\\
\omega_{1,to}^{z} &= 135\unit{cm}^{-1} & &= 4.1\unit{THz}\,,
\label{eqn:omega_to}
\end{alignedat}
\end{equation}
differ more than twice.
As a result, this and similar TIs can exhibit a giant anisotropy of the dielectric permittivity.
There is a range of $\omega$ where the permittivity tensor is indefinite:
the real part of $\epsilon^z(\omega)$ is positive, while that of $\epsilon^\bot(\omega)$ is negative.
Media with such characteristics
are referred to as hyperbolic~\cite{Guo2012, Poddubny2013, Sun2014} because 
the isofrequency surfaces of their extraordinary rays
in the momentum space $\mathbf{k} = (k^x, k^y, k^z)$ are shaped as hyperboloids [Fig.~\ref{fig:model_b}(a)].
In the THz domain,
the widest band of frequencies where \BiSe*
behaves as a hyperbolic medium (HM) is
between the aforementioned dominant frequencies,
$\omega_{to, 1}^\bot < \omega < \omega_{to, 1}^z$;
however, other hyperbolic bands also exist in this TI
(both at THz frequencies, see Sec.~\ref{sec:model},
and at visible frequencies, see Ref.~\onlinecite{Esslinger2014}).
It is important that the approximate equation for the extraordinary isofrequency surfaces,
\begin{equation}
\frac{(k^x)^2 + (k^y)^2}{\epsilon^z(\omega)} +
\frac{(k^z)^2}{\epsilon^\bot(\omega)} =
\frac{\omega^2}{c^2}\,,
\label{eqn:HP2_bulk_dispersion}
\end{equation}
is valid up to $|\mathbf{k}|$
of the order of the inverse lattice constant.
Accordingly, rays of momenta $|\mathbf{k}|$ greatly exceeding the free-space photon momentum $\omega / c$ can propagate through hyperbolic materials without evanescent decay.
At such $\mathbf{k}$
the hyperboloids can be further approximated by cones,
which means that
the group velocity $\mathbf{v} = \partial\omega / \partial{\mathbf{k}}$ of the rays makes a \textit{fixed} angle $\theta$ (or $-\theta$) with respect to the $z$-axis, with
\begin{equation}
\tan\theta(\omega) = i\,
 \frac{[\epsilon^\bot(\omega)]^{1 / 2}}
      {[\epsilon^z(\omega)]^{1 / 2}}
      \,,
\label{eqn:theta}
\end{equation}
see Fig.~\ref{fig:model_b}(a). We refer to these deeply subdiffractional, highly directional modes as the hyperbolic phonon polaritons (HPP or HP$^2$, for short).

Our interest to HP$^2$ of TIs is stimulated by recent
discovery~\cite{Dai07032014, Jacob2014npp} and further exploration of similar collective modes in other systems such as
hexagonal boron nitride~\cite{Caldwell2014, Dai2015sfg, Li2015hpp, Shi2015apr} (hBN)
and hBN covered by graphene~\cite{Brar2013hct, Dai2015goh, Ni2015} (hBN/G).
There is a close analogy between these systems.
In fact, except for the difference in the number of Dirac cones
($N = 1$ \textit{vs}. $N = 4$)
and the frequency range where the hyperbolic response occurs (THz \textit{vs}. mid-infrared),
the electrodynamics of longitudinal collective modes of \BiSe* and hBN/G structures is qualitatively the same.
(The analogy is the most faithful when graphene and hBN
are rotationally misaligned;
otherwise,
their collective modes are modified by the moir\'e superlattice effects.~\cite{Tomadin2014gmp, Ni2015})

\begin{figure}[b]
\includegraphics{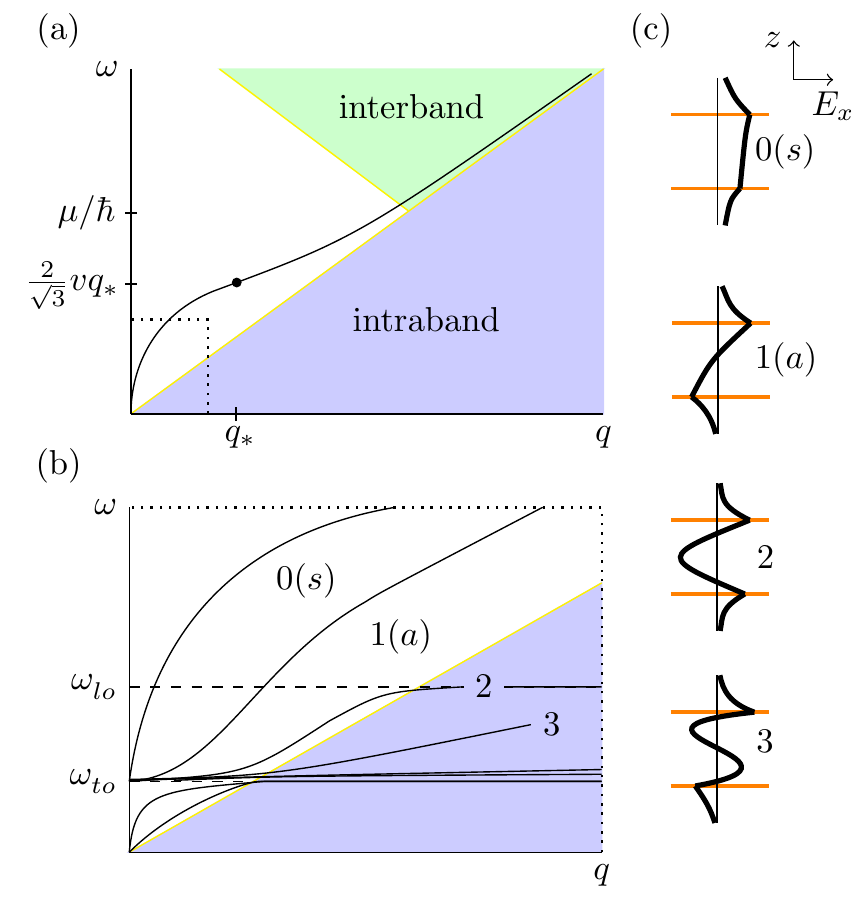}
  \caption{(Color online)
Schematic illustrations of the collective mode spectra in idealized model systems.
(a) The plasmon dispersion of Dirac fermions confined to the interface of two bulk media of constant positive permittivity
$\epsilon_0$ and $\epsilon_s$.
The dispersion crosses over from $\omega \simeq v\sqrt{q q_* / 2}$ to $\omega \simeq v q$
at a characteristic momentum $q_*$ [Eq.~\eqref{eqn:omega_1}].
The shaded areas indicate the electron-hole continua where the plasmons (and any other charged collective modes) are damped.
(b) The dispersion of hybrid HP$^3$ modes for a slab of a hypothetical TI material that has a single in-plane phonon mode at $\omega_{to}$ and constant $\epsilon^z > 0$. Permittivity $\epsilon^\bot$ is negative at $\omega_{to} < \omega < \omega_{lo}$ and positive at other $\omega$.
The dotted boundary corresponds to the dotted line in (a).
Outside the band $\omega_{to} < \omega < \omega_{lo}$,
only plasmonic modes $0$ and $1$ exist.
In the degenerate case $\epsilon_0 = \epsilon_s$ they correspond to the symmetric ($s$)
and antisymmetric ($a$) combinations of the top and bottom interface plasmons.
Inside that band, multiple branches of HP$^3$ are formed due to hybridization of the plasmons with the HP$^2$ waveguide modes.
The frequencies of all the branches other than $0$ and $1$ tend to $\omega_{lo}$ at large momenta.
(c) Schematic in-plane electric field profiles of the first few HP$^3$ modes (thick curves).
The number of nodes in each profile (the points where they cross with the vertical lines $E_x = 0$) is equal to the modal index.
}\label{fig:rp_slab_disp}
\end{figure}

The main goal of this paper is to investigate the interaction of HP$^2$ with the Dirac plasmons of the topological SS.
The latter dominate
the charge (and current) density response of the system
at frequencies
outside the hyperbolic band
where HP$^2$ are absent.
Dirac plasmons have been extensively studied in previous literature~\cite{Castro2009, Hwang2009pms, Raghu2010, Fei2011, Fei2012, Chen2012, Grigorenko2012gp, Profumo2012, GarciadeAbajo2014gpc, Basov2014, DiPietro2013, Stauber2013, Schutky2013spd, Qi2014spp, Li2014tts, Stauber2014pds, Autore2015omb, Autore2015ppi} on both TI and graphene. 
The basic properties of the Dirac plasmons can be introduced
on the example of a hypothetical TI material
with a frequency-independent permittivity $\epsilon^z > 0$ and
the permittivity $\epsilon^\bot(\omega)$ dominated by a single phonon mode.
Such an idealized material
is hyperbolic in a single frequency interval
$\omega_{to} < \omega < \omega_{lo}$ where $\epsilon^\bot(\omega) < 0$.
Its Dirac plasmons exist at
$\omega < \omega_{to}$ and $\omega > \omega_{lo}$
where $\epsilon^\bot(\omega) > 0$.
In the setup shown in Fig.~\ref{fig:model_b}(b),
where the TI slab borders media of constant permittivities $\epsilon_0 > 0$ and $\epsilon_s > 0$,
there are two plasmon modes.
At large enough in-plane momenta
$q \equiv [(k^x)^2 + (k^y)^2]^{1 / 2}$
these modes are
confined to the opposite interfaces and electromagnetically decoupled.
In the relevant range of momenta $q < q_*$,
the dispersion of the plasmon bound to the top interface is given by
\begin{equation}
q(\omega) \simeq
\frac{4}{N}\, 
\frac{\epsilon_0 + \epsilon_1}{e^2 |\mu|} 
(\hbar\omega)^2\,,
\quad \hbar\omega \ll |\mu|\,,
\label{eqn:disp_drude}
\end{equation}
where
\begin{equation}
\epsilon_1(\omega) = [\epsilon^\bot(\omega)]^{1 / 2}\,
  [\epsilon^z(\omega)]^{1 / 2},
\label{eqn:epsilon_1}
\end{equation}
is the effective permittivity of the TI
and $\mu$ is the chemical potential of the SSs measured from the Dirac point.
At frequencies far below $\omega_{to}$ or far above $\omega_{lo}$,
function $\epsilon_1(\omega)$ can be approximated by a real constant,
which yields $\omega \propto \sqrt{q}$.
This typical two-dimensional (2D) plasmon dispersion
describes the low-frequency part of the full
curve sketched in Fig.~\ref{fig:rp_slab_disp}(a).
The plasmon dispersion for the bottom interface is obtained by replacing $\epsilon_0$ with $\epsilon_s$
(unless $\epsilon_s \gg \epsilon_0$,
in which case the range
$q > q_*$ is relevant where the dispersion is approximately linear, see Sec.~\ref{sub:surface}).

Equation~\eqref{eqn:disp_drude} implies
that the nature of the plasmon modes should change drastically
when $\omega$ enters the hyperbolic frequency band
where $\epsilon_1(\omega)$ [Eq.~\eqref{eqn:epsilon_1}] is imaginary and strongly $\omega$-dependent. 
This equation predicts a complex $q$,
which suggests that the Dirac plasmons become leaky modes that rapidly decay into the HP$^2$ bulk continuum.
However, this is not quite correct.
We will show that nonleaky, i.e., propagating modes can survive in thin enough
TI slabs where the HP$^2$ continuum is broken into discrete subbands of \textit{waveguide} modes.
The latter hybridize with plasmons to form hyperbolic plasmon phonon polaritons (HPPP or HP$^3$, for short),
the primary target of our investigation,
see Figs.~\ref{fig:rp_slab_disp}(b) and (c).
We explore
the following properties and manifestations of the collective charge modes of the TIs:
i) the mode dispersion in the momentum-frequency space,
ii) the dependence of such dispersions on the surface doping and the thickness of the slab,
iii) the unusual real-space dynamics of the HP$^3$ rays,
including a polaritonic analog of the Goos-H\"anchen (GH) effect.~\cite{Goos1947, Bliokh2013}

The remainder of the paper is organized as follows.
In Sec.~\ref{sec:model} we specify the model and the basic equations.
In Sec.~\ref{sec:result} we present our results for the dispersion of the three different types of collective modes (plasmons, HP$^2$s, and HP$^3$s).
In Sec.~\ref{sec:GH},
which is the centerpiece of this work,
we discuss
waveguiding and launching of the HP$^2$ modes and also their tunable GH shifts.
We explain how these phenomena can be probed experimentally
using the imaging capabilities of the scattering-type scanning near-field optical microscopy (s-SNOM).~\cite{Keilmann2004nfm, Atkin2012noi}
In Sec.~\ref{sec:conclusions} we give concluding remarks and an outlook for the future.
Finally,
in Appendix we discuss signatures of the phonon-plasmon coupling measurable by the s-SNOM operating in the spectroscopic mode.

\section{Model}
\label{sec:model}

\begin{figure}
\includegraphics[width=2.8in]{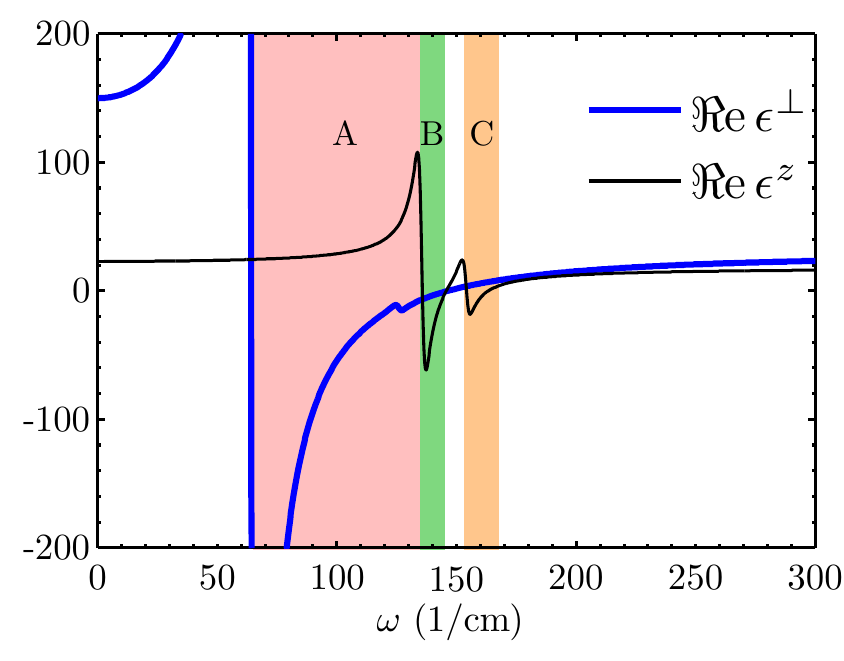}
\caption{(Color online) The real parts of the tangential and axial permittivities of Bi$_2$Se$_3$.
The sign changes of the permittivities are due to the $E_u$ and $A_{2u}$ phonons.
Surface- and bulk-confined collective modes exist inside the spectral regions where at least one of the permittivitties is negative. They include
type II hyperbolic region A ($\Re\mathrm{e}\,\epsilon^\bot < 0$, $\Re\mathrm{e}\,\epsilon^z > 0$),  Reststrahlen region B ($\Re\mathrm{e}\,\epsilon^\bot, \Re\mathrm{e}\,\epsilon^z < 0$), and type I hyperbolic region C ($\Re\mathrm{e}\,\epsilon^\bot > 0$, $\Re\mathrm{e}\,\epsilon^z < 0$).
}
\label{fig:epsi}
\end{figure}

Our model for the bulk permittivities of the TI is
\begin{equation}
\epsilon^\alpha(\omega)
= \epsilon_{\infty}^\alpha +
\sum_{j = 1, 2} \frac{\omega_{p, j}^{\alpha\,2}}
     {\omega_{to, j}^{\alpha\,2} - \omega^2 - i \gamma_j^{\alpha}\omega}
\,,
\quad
\alpha = \bot, z\,. 
\label{eqn:w_i_2}
\end{equation}
In the case of \BiSe*,
we choose the parameters based on available experimental~\cite{Richter1977, LaForge2010, DiPietro2012ocb}
and theoretical~\cite{Cheng2011} literature as follows:
$\epsilon_{\infty}^\bot = 29$,
$\epsilon_{\infty}^{z} = 17.4$,
$\omega_{to, 1}^{\bot} = 64\unit{cm}^{-1}$,
$\omega_{p, 1}^{\bot} = 704\unit{~cm}^{-1}$,
$\omega_{to, 2}^{\bot} = 125\unit{cm}^{-1}$,
$\omega_{p, 2}^{\bot} = 55\unit{cm}^{-1}$,
$\omega_{to, 1}^{z} = 135\unit{cm}^{-1}$,
$\omega_{p, 1}^{z} = 283\unit{cm}^{-1}$,
$\omega_{to, 2}^{z} = 154\unit{cm}^{-1}$, 
$\omega_{p, 2}^{z} = 156\unit{cm}^{-1}$,
and $\gamma_j^{\alpha} = 3.5\unit{cm}^{-1}$.
[Note that $\omega_{to, 1}^{\bot}$ and $\omega_{to, 1}^{z}$ were already listed in Eq.~\eqref{eqn:omega_to}.]
The real parts of functions $\epsilon^\bot(\omega)$ and $\epsilon^z(\omega)$ are plotted in Fig.~\ref{fig:epsi}.
The regions where at least one of them is negative are shaded.
They include
region A, $\omega_{to, 1}^\bot < \omega < \omega_{to, 1}^z$, where \BiSe* is a HM of type II
($\Re\mathrm{e}\,\epsilon_z > 0$, $\Re\mathrm{e}\,\epsilon^\bot < 0$);
region C, $\omega_{to, 2}^z < \omega < 163\unit{cm}^{-1}$
where it is a HM of type I ($\Re\mathrm{e}\,\epsilon_z < 0$, $\Re\mathrm{e}\,\epsilon^\bot > 0$),
and region B, $\omega_{to, 1}^z < \omega < 146\unit{cm}^{-1}$,
where it exhibits the Reststrahlen behavior ($\Re\mathrm{e}\,\epsilon_z < 0$, $\Re\mathrm{e}\,\epsilon^\bot < 0$).
Since regions B and C are narrow,
in our discussion of HP$^2$ and HP$^3$ modes we focus on region A.
In this discussion we often refer to hBN as an example of a simpler material.
The type II hyperbolic band of hBN is bounded by
the frequencies~\cite{Dai07032014, Caldwell2014}
\begin{equation}
\omega_{to} = 1376\unit{cm}^{-1},
\quad
\omega_{lo} = 1614\unit{cm}^{-1}.
\label{eqn:omega_hBN}
\end{equation}
In this band $\epsilon^\bot(\omega)$ of hBN can be modelled
similar to Eq.~\eqref{eqn:w_i_2} but using a single Lorentzian oscillator while
$\epsilon^z$ can be considered $\omega$-independent and positive.

In the case of \BiSe*, we
also have to specify our assumptions
about the electronic response.
We consider only frequencies smaller than the bulk gap $0.3\unit{eV}$ of \BiSe*
at which the electronic contribution to the permittivities [included in Eq.~\eqref{eqn:w_i_2} via $\epsilon_\infty^\alpha$] is purely real.
Additionally, we assume that the valence bulk band is completely filled,
the conduction one is empty,
with no free carriers present in the bulk.
However, such carriers populate the gapless SS
described by the massless 2D Dirac equation.
The chemical potential $\mu$,
which is located inside the bulk band gap,
determines the doping of these SS.
We ignore any virtual or real electronic transitions between the surface and the bulk states.

%

The fundamental current/density response functions of the SS are the sheet conductivity $\sigma$ and
polarizability $P$,
which are related in the standard way:
\begin{equation}
\sigma(q,\omega) = \frac{i \omega}{q^2}\,
 e^2 P(q,\omega)\,.
\label{eqn:sigma_p}
\end{equation}
Within the random-phase approximation for Dirac fermions,
$P(q,\omega)$ can be computed~\cite{Wunsch2006, Hwang2007} analytically:
\begin{equation}
\begin{split}
P(q, \omega) &= -\frac{N k_F}{2\pi\hbar v}
-
\frac{i N}{16\pi \hbar v}
 \frac{q^2}{\sqrt{q^2 - k_\omega^2}}
\\
&\times \left[G\left(\frac{k_\omega + 2 k_F}{q}\right)
- G\left(\frac{k_\omega - 2 k_F}{q}\right)
-i\pi\right],
\\
G(x) &= i x\sqrt{1 - x^2} - i \arccos x\,.
\end{split}
\label{eqn:p3}
\end{equation}
Here the branch cut for the square root and logarithm functions is 
the negative real semi-axis,
$k_\omega$ is defined by $k_\omega = (\omega + i \gamma_e) / v$, phenomenological parameter $\gamma_e > 0$
is the electron scattering rate,
$v$ is the Fermi velocity,
and $k_F = |\mu| / (\hbar v)$ is the Fermi momentum.
Equation~\eqref{eqn:p3} is a good approximation at small $\mu$.
At large doping, trigonal warping~\cite{LeBlanc2014dss} and other details of realistic band-structure~\cite{Li2014tts} should be included.
Since the above formula is a bit cumbersome,
it may be helpful to mention some properties of $\sigma(q, \omega)$.
For example, if $\gamma_e = +0$,
the real part of $\sigma(q, \omega)$
is nonvanishing only inside the
two shaded areas in Fig.~\ref{fig:rp_slab_disp}(a),
which together form the so-called electron-hole continuum.~\cite{Castro2009, Basov2014}
(This real part is a measure of dissipation, i.e., Landau damping.)
For doped system at small momenta and frequencies,
$q, k_\omega \ll k_F$,
the expression for the conductivity can be reduced to
\begin{equation}
\sigma(q, \omega) \simeq
\frac{N e^2}{2 \pi \hbar}\,
\frac{k_F}{\sqrt{q^2 - k_\omega^2}}\,
\frac{i k_\omega}{i k_\omega - \sqrt{q^2 - k_\omega^2}}
\,.
\label{eqn:sigma_doped}
\end{equation}
At $q \ll \omega / v$, it further simplifies to the Drude formula
\begin{equation}
\sigma \simeq
\frac{N e^2}{4 \pi \hbar^2}\,
\frac{|\mu|}{\gamma_e - i\omega}\,,
\quad \mu \neq 0\,.
\label{eqn:drude}
\end{equation}
For an undoped system, one finds instead
\begin{align}
\sigma(q, \omega) &=
\frac{N}{16}\, \frac{e^2}{\hbar}
\frac{i k_\omega}{\sqrt{q^2 - k_\omega^2}}
\label{eqn:sigma_undoped}\\
&\simeq
\frac{N}{16}\,
\frac{e^2}{\hbar}
\,,
\quad q \ll \frac{\omega}{v}\,.
\label{eqn:universal}
\end{align}

In order to find the dispersion of the collective modes of the TI slab
we use two computational methods.
One method,
which is advantageous for deriving analytical results,
is to
look for the poles of the response function $r_{P}(q, \omega)$.
This function is the total ${P}$- (also known as the $\mathrm{TM}$-) polarization reflectivity of the system
measured when an external field is incident from the medium labeled ``$\epsilon_0$'' in Fig.~\ref{fig:model_b}(b). 
It must be immediately clarified that
$r_{P}(q, \omega)$ has no poles at 
simultaneously real $q$ and $\omega$ if the dissipation parameters $\gamma$ and $\gamma_e$ are nonzero.
At least one of these arguments must be complex.
Whenever one refers to the dispersion relation of a mode,
one means
the relation between the real parts of $q$ and $\omega$.
The other method, which is especially convenient for numerical simulations,
is to identify the sought dispersion curves with the maxima of $\Im\mathrm{m}\, r_{P}(q, \omega)$
at \textit{real} arguments.
As long as the imaginary parts of $q$ and $\omega$
(which give information about the propagation length and lifetime of the mode) are small,
both methods give the same dispersions.
An extra benefit of working with real $q$ and $\omega$ 
is that the corresponding
$r_{P}(q, \omega)$ is the input for
further calculations we discuss in Appendix~\ref{sec:SNOM}
where we model s-SNOM experiments for the system in hand.

\begin{figure*}
\includegraphics[width=6.0in]{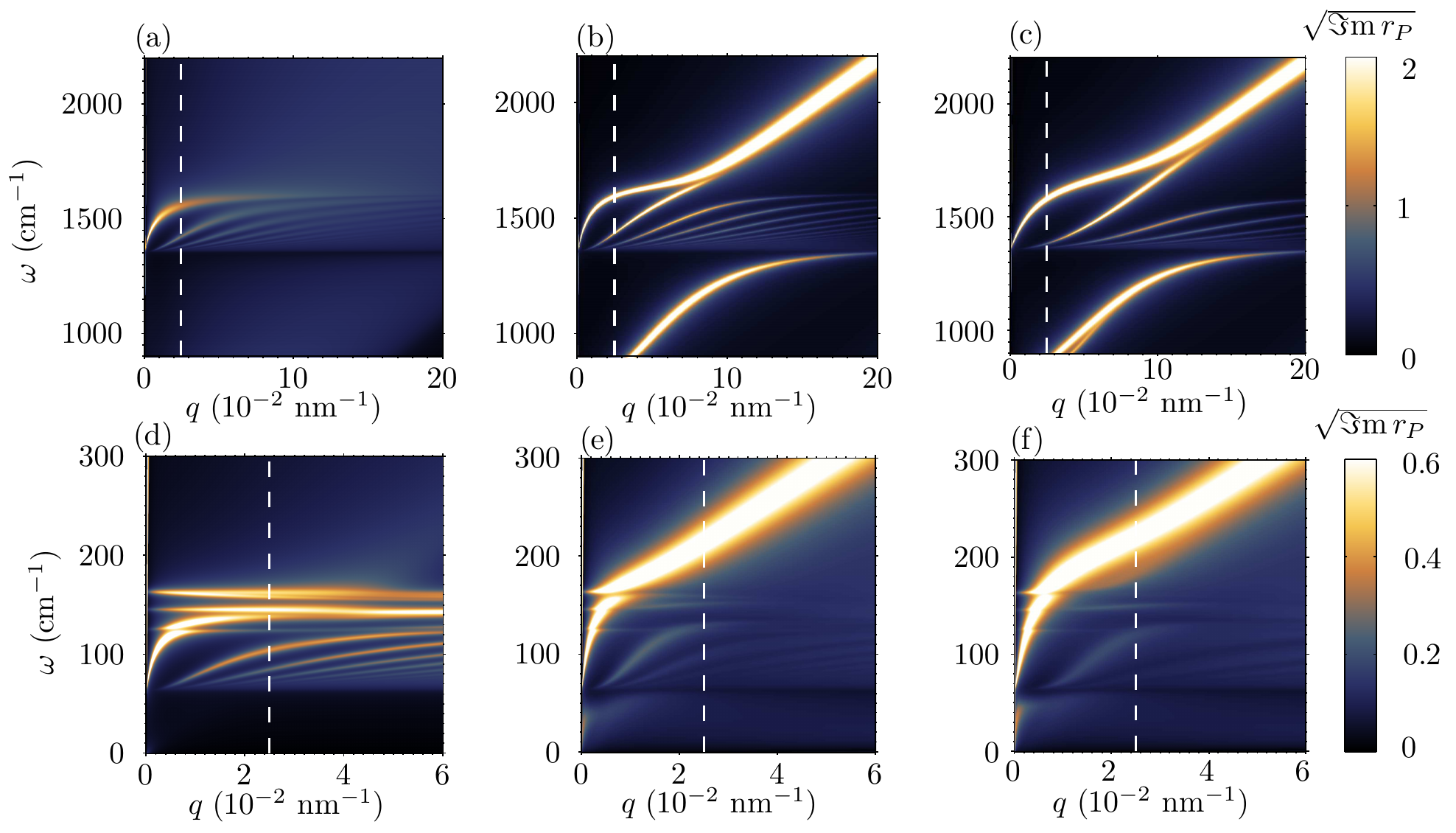}
\caption{(Color online)
Collective mode dispersions of graphene-hBN-graphene (\GHG*) and \BiSe* slabs rendered
using the false color maps of $\Im\mathrm{m}\, r_{P}$.
The parameters of the calculation  for \GHG* are:
(a) $d = 60\unit{nm}$, $\mu = 0$,
(b) $d = 60\unit{nm}$, $\mu = 0.29\unit{eV}$,
(c) $d = 30\unit{nm}$, $\mu = 0.29\unit{eV}$.
The other parameters are $v = 1.00\times 10^8\unit{cm/s}$,
$\gamma_e = 1.00\unit{THz}$, $\epsilon_0 = 1$, and
$\epsilon_s = 1.5$.
The parameters of the calculation for \BiSe* are:
(d) $d = 120\unit{nm}$, $\mu = 0$,
(e) $d = 120\unit{nm}$, $\mu = 0.29\unit{eV}$,
(f) $d = 60\unit{nm}$, $\mu = 0.29\unit{eV}$.
In these three plots $v = 0.623 \times 10^8\unit{cm/s}$,
$\gamma_e = 1\unit{THz}$, $\epsilon_0 = 1$, and
$\epsilon_s = 10$.
Equal doping of the top and bottom SS is assumed.
The vertical dashed lines indicate a characteristic momentum probed by the s-SNOM experiments simulated
in Fig.~\ref{fig:s3} below. 
}
\label{fig:rp_slab}
\end{figure*}

Our procedure for calculating function $r_{P}(q, \omega)$ can be explained as follows.
Taking a more general view for a moment,
we regard the entire system including the substrate and superstrate as a stack of layers $j = 0, 1, \ldots, M$ of thickness $d_j$,
tangential permittivity $\epsilon_{j}^\bot$, and axial permittivity $\epsilon_{j}^z$.
(In the present case, $M = 2$, the TI slab is layer $j = 1$ and $d_1 = d$.)
Additionally, we assume that the interface of the layers $j$ and $j + 1$ possesses the sheet conductivity $\sigma_{j, j + 1}$.
We observe that
the ${P}$-polarization reflectivity $r_{j, j + 1}$ of ${j, j + 1}$ interface in isolation is given by the formula (see, e.g., 
Ref.~\onlinecite{Dai2015goh})
\begin{gather}
r_{j,j+1} = \frac{Q_{j + 1} - Q_j + \dfrac{4\pi}{\omega} \sigma_{j,j+1}}
{Q_{j + 1} + Q_j + \dfrac{4\pi}{\omega} \sigma_{j,j+1}}\,,
\label{eqn:rp_unit}\\
Q_j = \frac{\epsilon_{j}^\bot}{k_{j}^z}\,,
\quad
k_{j}^z = \sqrt{\epsilon_{j}^\bot}\, \sqrt{\frac{\omega^2}{c^2} - \frac{q^2}{\epsilon_{j}^z}}
\,,
\label{eqn:k^z}
\end{gather}
where $k_{j}^z$ and $q$ are, respectively, the axial and the tangential momenta inside layer $j$.
Let $r_{j}$ be the reflectivity of a subsystem composed of layers $j,\ldots, M$.
By this definition, $r_{M - 1} = r_{M - 1, M}$.
The crucial point is that the desired $r_{P} \equiv r_0$ can be found by the backward recursion
\begin{equation}
r_j = r_{j, j+1}
- \frac{(1 - r_{j, j+1}) (1 - r_{j+1, j}) r_{j+1}}
{r_{j+1,j} r_{j+1} - \exp(-2 i k_{j+1} d_{j+1})}\,,
\label{eqn:rp_recursive}
\end{equation}
where $r_{j + 1, j}$ is the right-hand hand of Eq.~\eqref{eqn:rp_unit} with $Q_j$ and $Q_{j + 1}$ interchanged. 
For $M = 2$, one recursion step suffices,
which gives us, after some algebra,~\cite{Dai2015goh}
\begin{equation}
r_{P} = \frac{r_{1 2} (r_{0 1} + r_{1 0} - 1) - r_{0 1}
               \exp(-2 i k_1 d_1)}
             {r_{1 0} r_{1 2} - \exp(-2 i k_1 d_1)}\,.
\label{eqn:rp_recursive_II}
\end{equation}
Hence, function
$r_{P}(q,\omega)$ has poles whenever
\begin{equation}
r_{1 0}(q,\omega) r_{1 2}(q,\omega)
= \exp\left(-2 i k_1^z d \right).
\label{eqn:rp_pole}
\end{equation}
For large in-plane momenta $q \gg (\omega / c)
\max |\epsilon_j^z|^{1 / 2}$,
we can use the approximations
$k_1^z \simeq q \tan\theta$ and
\begin{gather}
r_{1 0} \simeq \frac{\epsilon_0 - \epsilon_1 - \frac{2 q}{q_{\mathrm{top}}}}
     {\epsilon_0 + \epsilon_1 - \frac{2 q}{q_{\mathrm{top}}}}
\,,
\quad
q_{\mathrm{top}}
\equiv \frac{i\omega}{2\pi \sigma_{\mathrm{top}}}\,,
\label{eqn:r_10}
\end{gather}
where $\sigma_{\mathrm{top}} = \sigma_{\mathrm{top}}(q, \omega)$
is the sheet conductivity of the SS at the top interface.
Let us also define the ``phase shifts''
$\phi_{\mathrm{top}}$ and $\phi_{\mathrm{bot}}$
for inner reflections from the top and bottom interfaces, respectively:
$r_{1 0} = -\exp(2 i \phi_{\mathrm{top}})$,
$r_{1 2} = -\exp(2 i \phi_{\mathrm{bot}})$.
Note that in general $\phi_{\mathrm{top}}$ and $\phi_{\mathrm{bot}}$ are complex numbers.
Specifically, we take
\begin{align}
\phi_{\mathrm{top}} &= \arctan
\left[i\,\frac{\epsilon_0}{\epsilon_1}
\left(1 - \dfrac{2}{\epsilon_0}
 \dfrac{q}{q_{\mathrm{top}}}\right)\right],
\label{eqn:phi_t}\\
\phi_{\mathrm{bot}} &= \arctan
\left[i\,\frac{\epsilon_s}{\epsilon_1}
\left(1 - \dfrac{2}{\epsilon_s}
 \dfrac{q}{q_{\mathrm{bot}}}\right)\right].
\label{eqn:phi_b}
\end{align}
where the standard definition of $\arctan z$ is assumed,
with the branch cuts $(-i\infty, -i)$, $(i, i \infty)$ in the complex-$z$ plane;
$q_{\mathrm{bot}}$ is defined analogously to
$q_{\mathrm{top}}$ but
with the sheet conductivity $\sigma_{\mathrm{bot}}$ of the bottom SS instead of $\sigma_{\mathrm{top}}$.
Equation~\eqref{eqn:rp_pole} can now be transformed to
\begin{equation}
q_n = -\frac{2}{\delta}(n \pi + \phi_{\mathrm{top}} + \phi_{\mathrm{bot}})
\,,
\quad
\delta \equiv 2 d \tan\theta\,,
\label{eqn:disp_film}
\end{equation}
where the integer subscript $n$ labels possible multiple solutions.
Admissible $n$ must satisfy the condition $\Im\mathrm{m}\, r_{P}(q_n, \omega) > 0$.
Our numerical results for $r_{P}$ computed from Eq.~\eqref{eqn:rp_recursive_II}
and analytic approximations
for the solutions of Eq.~\eqref{eqn:disp_film} are presented in Sec.~\ref{sec:result}.

\section{Collective mode dispersions}
\label{sec:result}

The false color maps of function $\Im\mathrm{m}\, r_{P}(q, \omega)$ provide a convenient visualization of
the collective mode spectra.
Examples of such maps computed for \BiSe* slabs are presented
in the bottom row of Fig.~\ref{fig:rp_slab}.
Their counterparts for graphene-hBN-graphene (\GHG*) structures are shown in the top row to facilitate the interpretation.
The bright lines in Fig.~\ref{fig:rp_slab} are the dispersion curves of the collective modes.
The apparent widths of those lines give an idea how damped the modes are.
Below we discuss these results in more detail.

\subsection{Hyperbolic waveguide modes}


Figures~\ref{fig:rp_slab}(a) and \ref{fig:rp_slab}(d)
depict the $\Im\mathrm{m}\, r_{P}$
maps for,
respectively, \GHG* and \BiSe* slabs,
when they are undoped, $\mu = 0$.
No Dirac plasmons exist in such systems,
so that the collective modes are limited to HP$^2$s.
In Fig.~\ref{fig:rp_slab}(a) we see a single family of 
such modes whereas in \ref{fig:rp_slab}(d) one can actually distinguish three of them.
Let us start with the former, simpler case.
The key to understanding the nature of these modes is that
inside the hyperbolic band $\omega_{to} < \omega < \omega_{lo}$ the $z$-axis momentum $k_1^z \simeq q \tan\theta$ of the modes is nearly real.
Hence, the HP$^2$s
form standing waves inside the slab.
The integer
$n$ in Eq.~\eqref{eqn:disp_film} corresponds to the number of nodes of these waves,
see Fig.~\ref{fig:rp_slab_disp}(c).
For \GHG* the requisite condition $\Im\mathrm{m}\, r_{P} > 0$ is
satisfied
by all nonegative integers $n$ due to the fact that
$\Im\mathrm{m}\,\tan\theta > 0$. 
This inequality also ensures that $\Im\mathrm{m}\, q > 0$.
An analytical approximation for the dispersion curves of an undoped 
slab is obtained by neglecting
the fractions $q / q_{\mathrm{top}}$, $q / q_{\mathrm{bot}}$ in
Eqs.~\eqref{eqn:phi_t}, \eqref{eqn:phi_b},
in which case
Eq.~\eqref{eqn:disp_film} yields $q(\omega)$ directly.
Within this approximation, momenta $q_n$ at given $\omega$ are equidistant:
\begin{equation}
q_{n + 1} - q_n \simeq -\frac{2\pi}{\delta}
= -\frac{\pi}{d}\,
 \frac{1}{\tan\theta(\omega)}\,.
\label{eqn:q_n_HP2}
\end{equation}
The dispersion of the HP$^2$ waveguide modes is dominated by the factor
$1 / \tan\theta(\omega)$ in Eqs.~\eqref{eqn:disp_film},
\eqref{eqn:q_n_HP2},
which, if all damping is neglected, changes from zero to infinity as $\omega$ increases from $\omega_{to}$ to $\omega_{lo}$.
This is precisely what we see in Fig.~\ref{fig:rp_slab}(a):
all the dispersion
curves start at $\omega_{to}$ at $q = 0$ and increase toward $\omega_{lo}$ at large $q$.

Equation~\eqref{eqn:q_n_HP2} is general
and it applies to \BiSe* as well.
The three families of collective modes
seen in Fig.~\ref{fig:rp_slab}(d),
belong to the spectral regions A, B, and C of Fig.~\ref{fig:epsi}.
In region A, which is the widest of the three,
we see a set of HP$^2$ modes very similar to those in
Fig.~\ref{fig:rp_slab}(a).
They start at $\omega_{to, 1} = 64\unit{cm}^{-1}$ at $q = 0$ and
monotonically increase toward $\omega_{to, 2} = 135\unit{cm}^{-1}$ at large $q$.
In region C, $154 < \omega\,(\mathrm{cm}^{-1}) < 163$,
we again find a family of HP$^2$ modes but this time with a negative dispersion.
This behavior is typical of type I HM ($\Re\mathrm{e}\,\epsilon^\bot > 0$, $\Re\mathrm{e}\,\epsilon^z < 0$).
The shape of the dispersion can be understood noticing that the real part of $1 / \tan \theta(\omega)$ is positive,
varying from $\infty$ to $0$ (if the phonon damping $\gamma_j^\alpha$ is neglected)
while admissible $n$ are now $n \leq 0$.
[In hBN, this type I behavior is also realized~\cite{Caldwell2014, Dai2015goh, Li2015hpp} but the corresponding frequency range
is below the axis cutoff in Fig.~\ref{fig:rp_slab}(a).]
Lastly, in region B,
$135 < \omega\,(\mathrm{cm}^{-1}) < 146$,
function $\tan \theta(\omega)$ is almost purely imaginary,
which implies that the collective modes do not form standing waves but are exponentially confined to the interfaces.
Also, there are only two such modes, $n = 0$ and $n = 1$.
In this respect these surface-bound HP$^2$ modes are similar to the Dirac plasmons, see Sec.~\ref{sec:Introduction} above and Sec.~\ref{sub:surface} below.
However, their dispersion is completely different from those of the plasmons, e.g.,
the dispersion of the upper ($n = 1$) mode has a negative slope,
see Fig.~\ref{fig:rp_slab}(d).
Similar collective excitations have been studied
in literature devoted to other systems, e.g., anisotropic superconductors,~\cite{Stinson2014}
which can be consulted for details and references.
Due to narrowness of regions B and C, some of the described features may be difficult to see in Fig.~\ref{fig:rp_slab}(a)
and probably challenging to observe in experiments.
For this reasons, we will mostly refrain from discussing regions B and C further.

One implication of Eq.~\eqref{eqn:q_n_HP2} is that the HP$^2$ dispersion is widely tunable:
the scaling law $q_n \propto d^{-1}$ provides a practical way to engineer a desired wavelength of the waveguide modes simply by tailoring the slab thickness $d$,
as has been previously demonstrated using hBN slabs.~\cite{Dai07032014}

\subsection{Surface plasmons}
\label{sub:surface}

Examples of the collective mode spectra at finite doping
are shown in Fig.~\ref{fig:rp_slab}(b, c) for \GHG* and \ref{fig:rp_slab}(e, f) for \BiSe*.
The spectra are dramatically different
inside and outside the
hyperbolic frequency bands.
A key to understanding this difference is again the value of the momentum $k_1^z \simeq q \tan\theta(\omega)$.
Outside the hyperbolic bands, it is almost purely imaginary,
and so the collective excitations are exponentially confined to the surfaces of the slab.
These surface modes are the Dirac plasmons
introduced in Sec.~\ref{sec:Introduction}.
Having in mind applications to near-field experiments,
we are particularly interested
in momenta $q$ of the order of a few times $10^5\unit{cm}^{-1}$,
i.e., the region nearby the dashed lines
$q = 0.0025\unit{nm}^{-1}$ in
Fig.~\ref{fig:rp_slab}.
If $\epsilon_1$ is real, there are at most two solutions of Eq.~\eqref{eqn:disp_film},
one for $n = 0$ and the other for $n = 1$.
However, the distinct $n = 1$ dispersion curves are
visible only in Fig.~\ref{fig:rp_slab}(b, c)
for \GHG*
and none of them is close enough to  
the range of $q$ we are interested in.
Therefore, we focus on the $n = 0$ branch.

The shape of the plasmon dispersion curves in TI slabs and double-layer graphene systems
was a subject of many previous theoretical studies~\cite{Hwang2009pms, Profumo2012, Stauber2013, Stauber2014pds, Li2014tts}
whose basic conclusions are reproduced by the following analysis.
To the right of the dashed lines
in Fig.~\ref{fig:rp_slab}(b, e) and for $d \sim 100\unit{nm}$,
the dimensionless product $2 k_1^z d = q \delta$
is typically large by absolute value and
almost purely imaginary.
This implies that the plasmons of the two interfaces are decoupled.
Taking into account that
$\epsilon_0 < \epsilon_s$ and
$q_{\mathrm{top}} = q_{\mathrm{bot}}$
in Fig.~\ref{fig:rp_slab},
one can show that
the dispersion of the $n = 0$ mode is
controlled by the properties of the top interface.
In the first approximation this dispersion can be obtained setting
$\phi_{\mathrm{top}} \to -i \infty$,
which yields
\begin{equation}
q_0 \approx \frac{\epsilon_0 + \epsilon_1}{2}\,
q_{\mathrm{top}}\,,
\quad q_0 \gg |\delta|^{-1}.
\label{eqn:q_0}
\end{equation}
For $\mu = 0$, momentum $q_{\mathrm{top}} = q_{\mathrm{top}}(q_0, \omega)$ is imaginary,
cf.~Eqs.~\eqref{eqn:universal} and \eqref{eqn:r_10}.
Hence, for real $\epsilon_1$,
Eq.~\eqref{eqn:q_0} has no real solutions: as already mentioned,
undoped SS do not support plasmons.
Indeed, Figs.~\ref{fig:rp_slab}(a) and (d) contain no bright lines
outside the hyperbolic bands.
On the other hand, if $\mu \neq 0$,
we can use Eq.~\eqref{eqn:drude}
to transform Eq.~\eqref{eqn:q_0} to Eq.~\eqref{eqn:disp_drude},
which
predicts a parabolic dispersion curve $\omega \propto \sqrt{q}$ if $\epsilon_1$ is constant.
Such parabolas are seen in the upper halves of 
Figs.~\ref{fig:rp_slab}(b, c) and (e, f) although they appear rectilinear because of the restricted range of $q$.

As smaller momenta Eq.~\eqref{eqn:q_0} no longer holds.
The correct approximation for the $n = 0$ mode is obtained
by setting the left-hand side of Eq.~\eqref{eqn:disp_film} to zero.
This yields $\phi_{\mathrm{top}} = -\phi_{\mathrm{bot}}$ and
\begin{equation}
q_0 \simeq \frac{\epsilon_0 + \epsilon_s}{2}\,
\frac{1}{q_{\mathrm{top}}^{-1} + q_{\mathrm{bot}}^{-1}}
\simeq
\frac{2}{N}\, 
\frac{\epsilon_0 + \epsilon_s}{e^2 |\mu|} 
(\hbar\omega)^2\,.
\label{eqn:mode_0_long}
\end{equation}
Thus, both the low-$q$ and high-$q$ parts of the $n = 0$ dispersion curve are parabolic but with different curvatures.
The crossover between these two parabolas
occurs via a rapid increase
of $\epsilon^\bot(\omega)$, and so, $\epsilon_1(\omega)$ at frequencies immediately above the hyperbolic bands.
It takes place at $\omega > 1614\unit{cm}^{-1}$ for \GHG* and $\omega > 163\unit{cm}^{-1}$ for \BiSe*,
which generates
the inflection points seen on the curves in, respectively, Fig.~\ref{fig:rp_slab}(b, c) and (e, f).

As indicated schematically in Fig.~\ref{fig:rp_slab_disp}(a),
at very large $q$ the plasmon dispersion should have another inflection point.
Using the more accurate Eq.~\eqref{eqn:sigma_doped} instead
of Eq.~\eqref{eqn:drude},
we find the following analytical result for the frequency
of the $n = 0$ mode as a function of $q$:
\begin{equation}
\omega(q) \simeq v\,
\frac{q + q_*}{\sqrt{1 + (2{q_*}/{q})}}\,,
\quad
q_* = \frac{2 e^2}{\hbar v}\,
\frac{N k_F}{\epsilon_0 + \epsilon_1}
\,.
\label{eqn:omega_1}
\end{equation}
This equation predicts a crossover from the parabolic to the linear dispersion $\omega \simeq v q$ above $q = q_*$.
However, this occurs far outside the plot range of 
Fig.~\ref{fig:rp_slab}.

Returning to Eq.~\eqref{eqn:mode_0_long},
we notice that it does not contain the bulk permittivities.
Hence, it should continue to hold for a range of $\omega$ inside the hyperbolic bands.
A physical picture of this mode [``$0(s)$'' in Fig.~\ref{fig:rp_slab_disp}(c)] is in-phase oscillations of the charges of both Dirac fermion layers,
i.e., the system behaving as a single 2D layer with the combined oscillator strength.
As $\omega$ decreases further into the hyperbolic bands,
the length scale $|\delta|$ increases. 
The strength of the inequality $q_0 |\delta| \ll 1$
and so the accuracy of Eq.~\eqref{eqn:mode_0_long} becomes progressively lower [in fact, Eq.~\eqref{eqn:mode_0_very_long} below gives a better approximation].
At $\omega = \omega_{to}$ for \GHG* and similarly,
at $\omega = \omega_{to, 1}^\bot$ for \BiSe*,
this inequality is violated completely,
which is consistent with the termination of these branches at $q = 0$ in Figs.~\ref{fig:rp_slab}(b) and (e).
Similar analysis can be applied to Figs.~\ref{fig:rp_slab}(c) and (f) where $d$ is twice smaller than in, respectively,
Figs.~\ref{fig:rp_slab}(b) and (e).
Because of that, the plasmon dispersion in the region $q |\delta| < 1$ is shifted to smaller $q$.
The dispersions in the large-$q$ regions are virtually unaffected since the stronger surface confinement of the plasmons makes them insensitive to $d$.

One qualitative difference between \GHG* and \BiSe* is the
richer phonon spectrum of the latter.
This leads to the avoided crossings of the plasmon branch with the dispersion lines of the HP$^2$ modes in regions B and C of \BiSe*,
cf.~Fig.~\ref{fig:rp_slab}(b, c) and \ref{fig:rp_slab}(e, f).
The small shifts caused by those crossings are somewhat masked by the considerable linewidth of the $n = 0$ line
due to relatively stronger phonon damping.
In turn, higher electronic damping rate
$\gamma_e \sim \omega_{to, 1}^\bot$
due to disorder scattering in \BiSe*
effectively eliminates the plasmon excitations
in the lower spectral region $\omega < \omega_{to, 1}^\bot$,
see Fig.~\ref{fig:rp_slab}(e, f).
Therefore, we do not discuss it here.


\subsection{Hybrid modes}
\label{sub:HP3}

From now on we turn to the subject of our primary interest,
the hyperbolic collective modes of
a doped TI.
In this short section we address their dispersion law.
Comparing Fig.~\ref{fig:rp_slab}(d) for $\mu = 0$ with 
Fig.~\ref{fig:rp_slab}(e, f) for $\mu > 0$,
we observe significant shifts in the dispersion
of the $n = 0$ mode
in the upper half of the hyperbolic band
$\omega_{to, 1}^\bot < \omega < \omega_{to, 1}^z$ of \BiSe*.
Similar shifts are seen in hBN near
$\omega_{lo}$, cf.
Fig.~\ref{fig:rp_slab}(a) with 
Fig.~\ref{fig:rp_slab}(b, c).
These shifts result from hybridization of HP$^2$ and Dirac plasmons into combined HP$^3$ waveguide modes.
In general, calculation of these shifts requires solving Eq.~\eqref{eqn:disp_film} numerically.
However,
near the bottom of the hyperbolic band
where these shifts become small,
they can be also found analytically.
Thus, Eq.~\eqref{eqn:mode_0_long} gets replaced by
\begin{equation}
q_0 \simeq \dfrac{\epsilon_0 + \epsilon_s}
{\epsilon^\bot d
 + 2 q_{\mathrm{top}}^{-1} + 2 q_{\mathrm{bot}}^{-1}
 }
\,,
\quad |\epsilon_1| \gg \epsilon_0, \epsilon_s
\,,
\label{eqn:mode_0_very_long}
\end{equation}
which shows explicitly that $q_0$ goes to zero as $\omega$ approaches $\omega_{to, 1}^\bot$ where $\epsilon^\bot$ sharply increases.

Unlike in Fig.~\ref{fig:rp_slab}(a, d),
in \ref{fig:rp_slab}(b, c, e, f)
the higher-order $n > 1$ modes are more difficult to see
because of their lower relative intensity compared to
those of the plasmon $n = 0$ (and $n = 1$) modes. 
Nevertheless, these modes remain well defined (underdamped).
Near the bottoms of the respective hyperbolic bands their momenta $q_n$ still form an equidistant sequence except with a spacing
\begin{equation}
q_{n + 1} - q_n \simeq \frac{2\pi}{l - \delta}\,,
\label{eqn:q_n}
\end{equation}
which is modified compared to Eq.~\eqref{eqn:q_n_HP2}.
This result can be obtained
from Eq.~\eqref{eqn:disp_film} by
approximating the finite differences such as
$\phi_{\mathrm{top}}(q_{n + 1}) - \phi_{\mathrm{top}}(q_n)$
by means of the derivative.
Parameter $l$ is defined by
\begin{align}
l = -2\,
\frac{\partial \phi_{\mathrm{top}}}{\partial q}
- 2\,
\frac{\partial \phi_{\mathrm{bot}}}{\partial q}\,.
\label{eqn:gh_I}
\end{align}
The physical meaning of this quantity is clarified in the next Section.

\section{Goos-H\"anchen effect}
\label{sec:GH}

\begin{figure}[b]
\includegraphics[width=3.2in]{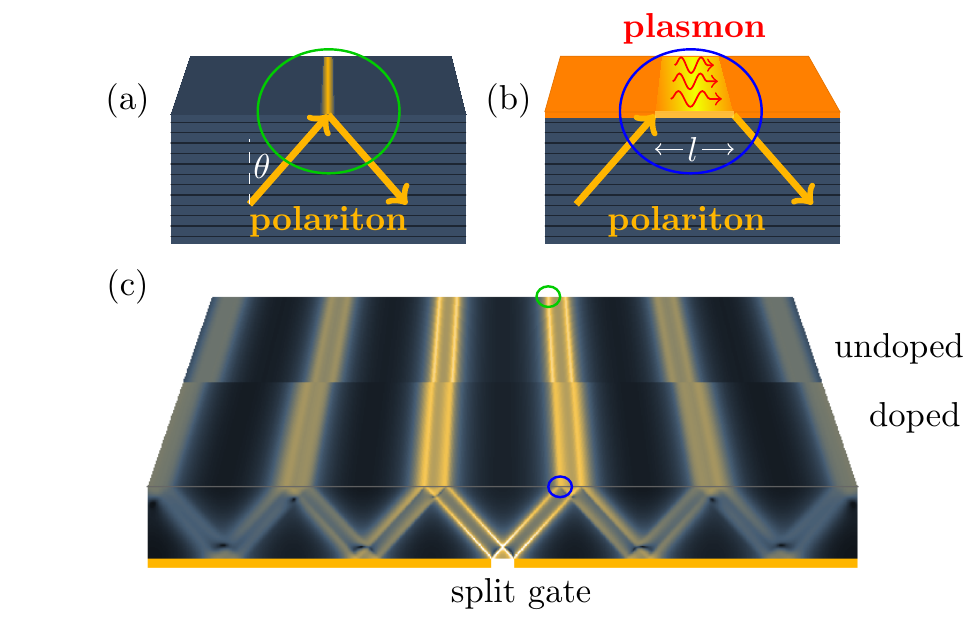}
  \caption{(Color online)
Polaritonic GH effect in TI slabs.
(a) Schematics of the HP$^2$ ray reflection in the absence of the SS.
(b) The same with the SS. The wavy lines symbolize virtual Dirac plasmons. The GH shift $l$ is indicated.  
(c) The electric field distribution inside and/or at the upper surfaces of two slabs with equal $\delta = -2.2 d$ but
different doping.
The lower 
(``doped'') and the upper (``undoped'')
parts of the image 
are computed for
$\lambda_p = a$ and $0$, respectively.
The split gate --- a pair of metallic half-planes separated by a distance $2a$ --- 
launches highly directional HP$^2$ rays that bounce inside the slabs creating periodic ``hot stripes'' at their upper surfaces.
The period is larger
in the ``doped'' slab.
The two small circles, one in the undoped and one in doped part, are the representative locations of the HP$^2$ reflections.
Their enlarged views are shown in, respectively, (a) and
(b).
}
\label{fig:gh}
\end{figure}

In this Section we consider the problem of the plasmon-polariton mixing from the point of view of real-space trajectories of the HP$^2$ excitations.
The question we consider is how polariton wavepackets 
propagate inside the slab and, in particular, how they
reflect off its interfaces.
As mentioned in Sec.~\ref{sec:Introduction},
for a given $\omega$,
the angle $\theta$ between the $z$-axis and the group velocity $\mathbf{v}$ vector of HP$^2$s 
is nearly independent of $q$.
Therefore, monochromatic HP$^2$ wavepackets propagate as highly directional rays.
Naively, one would then expect that
the polariton rays should zigzag inside the slab returning to each interface periodically with the repeat distance of $2 d \,|\!\tan\theta| = |\delta|$.
Although such geometrical optics picture
is adequate for insulating hyperbolic materials,~\cite{Sun2015hoh}
it is not quite correct for TI with gapless doped SS.
The geometrical optics
neglects a lateral shift or displacement of the rays after each reflection [compare Figs.~\ref{fig:gh}(a) and (b)],
which is analogous to the GH effect of light.
To define such a displacement one usually considers a wavepacket with a smooth envelope (for example, a Gaussian),
in which case
the displacement is the shift in the position of its maximum.

While the GH effect~\cite{Goos1947} was discovered measuring the reflection of light off an air-metal interface,
the displacement $\mathbf{l}$ of the reflected ray is a general wave phenomenon~\cite{Bliokh2013} that arises due to the dependence of the reflection phase shift $\phi$ on the lateral momentum $\mathbf{q} = (k^x, k^y)$.
For example, the GH effect should also occur for surface plasmons.~\cite{Huerkamp2011}
The expression for $\mathbf{l}$ has the form~\cite{Artmann1948}
\begin{align}
\mathbf{l} = -\Re\mathrm{e}\, \frac{\partial \phi}{\partial \mathbf{q}}\,.
\label{eqn:gh}
\end{align}
It seems to be another general rule that
the momentum dependence of $\phi$ is significant only if the interface supports electromagnetic modes with either a large propagation length or a long decay length if such modes are evanescent.
In the original photonic GH effect
this is the case under the conditions of the total internal reflection.
The magnitude $|\mathbf{l}|$ of the GH displacement
can be interpreted as the decay length of the evanescent transmitted wave.
Alternatively, a large GH shift can occur if the interface supports surface plasmons or polaritons.~\cite{Tamir1963sew, Tamir1971ldo, Chuang1986lso}
Experimental demonstration of the GH effect enhanced by surface plasmons of the air-metal interface has been reported.~\cite{yin2004} 

Comparing  Eqs.~\eqref{eqn:gh_I} and \eqref{eqn:gh},
we recognize the length scale $l$ in the former as the sum of the GH shifts due to the top and the bottom interfaces.
Therefore, we conclude that the Dirac plasmons must act
as the transient interface modes for the HP$^2$ rays bouncing inside the TI slab.
Using Eqs.~\eqref{eqn:phi_t}, \eqref{eqn:gh}, and taking into account
that $\Re\mathrm{e}\, \epsilon_1 \ll \Im\mathrm{m}\, \epsilon_1$,
we find the GH shift at the top interface to be
\begin{equation}
l_{\mathrm{top}} = \frac{4}{q_{\mathrm{top}}}
\,
\frac{\Im\mathrm{m}\, \epsilon_1}
{\bigl(\epsilon_0 - \frac{2 q}{q_{\mathrm{top}}}\bigr)^2 + |\epsilon_1|^2}\,.
\label{eqn:l}
\end{equation}
A few comments on this result can be made.
First, the GH shift is positive in our case,
which means the displacement is in the same direction as the in-plane group velocity of the ray.
Second, $l_{\mathrm{top}}$ depends on the permittivity of the environment.
For example,
at fixed $q$,
it vanishes if $\epsilon_0$ is very large.
Conversely, for fixed $\epsilon_0$, the GH shift 
reaches its maximum
\begin{equation}
l_\mathrm{{max}} = \frac{2}{\pi}\,
\frac{\lambda_p \epsilon_0 \Im\mathrm{m}\, \epsilon_1}
{\left(\Re\mathrm{e}\, \epsilon_1\right)^2
+ \left(\Im\mathrm{m}\, \epsilon_1\right)^2}
\,,
\quad
\lambda_p \equiv \frac{2\pi}{\epsilon_0 q_{\mathrm{top}}}\,,
\label{eqn:l_max}
\end{equation}
at $q = \pi / \lambda_p$.
Finally, $l_\mathrm{{max}}$ depends linearly on
the characteristic size $\lambda_p$ of the Dirac plasmon wavelength and inversely on the absolute value $|\epsilon_1| \approx \Im\mathrm{m}\, \epsilon_1$
of the effective permittivity of the hyperbolic medium.

\begin{figure}[t]
\includegraphics[width=3in]{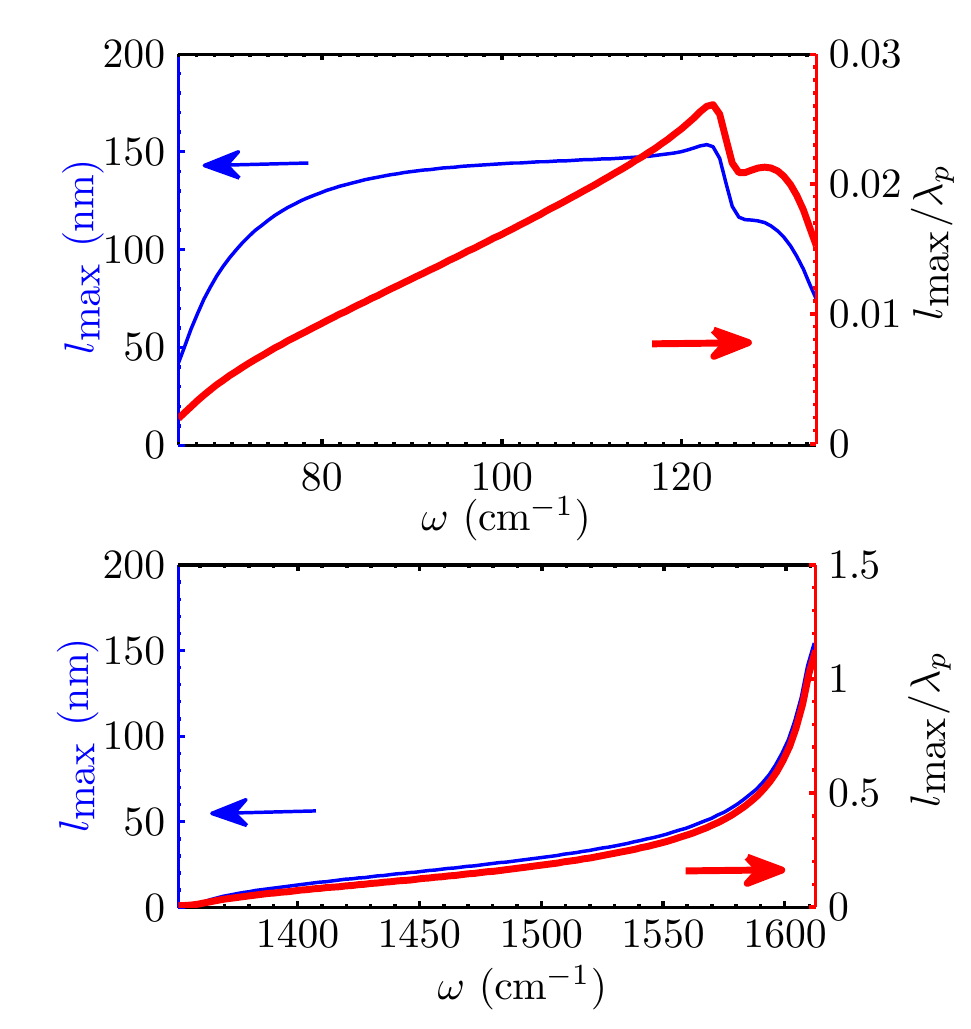}
\caption{(Color online)
Maximum GH shift $l_{\textrm{max}}$ (in absolute units and as a fraction of $\lambda_p$) for
(a) TI slab and (b) \GHG* structure
with the same chemical potential $\mu = 0.3\unit{eV}$.
}
\label{fig:gh2}
\end{figure}

In Fig.~\ref{fig:gh2}, we show $l_\textrm{max}$ for \BiSe* and \GHG* systems as a function of $\omega$ spanning their respective hyperbolic bands.
The relative shift, $l_\textrm{max}/\lambda_p$, is greater in \GHG* because $|\epsilon_1|$ is smaller.
Yet the absolute $l_\mathrm{max}$ at the same $\mu = 0.3\unit{eV}$ is greater in \BiSe* (where it is $\sim 200\unit{nm}$) because
it is hyperbolic
at lower frequencies and
$\lambda_p$ is larger at smaller $\omega$.

One possible setup for experimental detection of the GH effect in TI is shown in Fig.~\ref{fig:gh}(c).
It differs from Fig.~\ref{fig:model_b}(b) in the addition of a split gate between the TI slab and the substrate.
If this gate is made of a good conductor with
large permittivity, it would suppress the GH shift at the bottom surface.
However, it would serve another useful purpose.
Previously, it has been demonstrated~\cite{Dai2015sfg} that
in the presence of an external oscillating field,
thin metallic disks or stripes can launch HP$^2$
in hBN.
The split gate is to perform the same function here.
The HP$^2$s are preferentially emitted from the regions of highly concentrated field near the sharp metallic edges.
We expect the rays to zigzag away from their launching points
returning to the top surface with the period
$l -\delta$, which is the sum of
$-\delta \approx |\delta|$
due to the roundtrip inside the slab and $l = l_{\mathrm{top}}$ due to the GH shift at the top surface. 
Since $l$ depends $q_{\mathrm{top}}$, which is controlled by doping,
the GH effect can be detected by measuring the positions of the electric field maxima [``hot stripes'' in Fig.~\ref{fig:gh}(c)] as a function of $\mu$ in the experiment.
Although $l$ is quite small, the shifts accumulate after multiple reflections,
which can facilitate their detection,
as in the original work of Goos and H\"anchen.~\cite{Goos1947}

To model the response of the system
shown in Fig.~\ref{fig:gh}(c) quantitatively
we proceed as follows.
We approximate the half-planes of the split gate by perfect conductors in the $z = 0$ plane with the edges at $x = \pm a$.
Let $V(x, 0)$ be the scalar potential at $z = 0$ due to the external uniform field and all the charges induced on the gate.
(Here and below the common factor $e^{-i \omega t}$ is omitted.)
Let $\widetilde{V}(k^x)$ be the Fourier transform of $V(x, 0)$.
Using the notations for the reflection coefficients introduced in Sec.~\ref{sec:model},
we express the potential $V(x, z)$ inside the slab $0 \leq z \leq d$
by the integral
\begin{align}
V(x, z) &= \int \frac{d k^x}{2 \pi}\,
\widetilde{V}(k^x) t(k^x)
e^{i k^x x + i |k^x| z \tan\theta},
\label{eqn:V}\\
t(k^x) &= \frac{1 - r_{1 0}(k^x) r_{1 2}(k^x)}
{1 - r_{1 0}(k^x) r_{1 2}(k^x) e^{i k^x \delta} }\,.
\end{align}
For a consistency check we
can consider the large-$x$ behavior of this inverse Fourier transform, which should be dictated by the poles of the integrand.
These poles can be recognized as the HP$^3$ momenta $q_n$ [Eq.~\eqref{eqn:disp_film}].
Since $q_n$ form the equidistant sequence [Eq.~\eqref{eqn:q_n}],
their superposition should indeed create beats of period $l - \delta$,
in agreement with our ray trajectories picture,
Fig.~\ref{fig:gh}(b).

Explicit calculation of $V(x, 0)$
requires a self-consistent solution of the Maxwell equations for our complicated multilayer system,
which is computationally intensive.
Fortunately, very similar results for $V(x, z)$ are obtained with little effort by approximating the true $V(x, 0)$ with the ``bare'' potential that would exist
in the TI is removed, that is, if $d = \lambda_p = 0$.
At distances less than $c / \omega$ from the gap in the gate,
this bare potential has the simple analytical form,
\begin{equation}
V(x, 0) = \frac{V_0}{2} \times
\left\{\begin{alignedat}{2}
& {+1}\,,                     && x \leq -a\,,\\ 
& {-\frac{2}{\pi}} \arcsin ({x} / {a}),\quad & |& x| < a\,,\\
& {-1}\,,                      && x \geq a\,,
\end{alignedat}
\right.
\label{eq:mode_0_1}
\end{equation}
familiar from classical electrostatics.
Its Fourier transform is given by
\begin{equation}
\widetilde{V}(k^x) = \frac{i V_0}{k^x}\, J_0(k^x a)\,,
\label{eqn:V_launch}
\end{equation}
where $J_0(x)$ is the Bessel function of the first kind and
$V_0$ is potential difference between the two parts of the gate.
The tangential electric field corresponding to this potential,
\begin{equation}
E_x = \frac{V_0}{\pi \sqrt{a^2 - x^2}},
\end{equation}
exhibits an inverse square-root divergence at the edges,
which enables the localized HP$^2$ emission.
 
Carrying out the quadrature in Eq.~\eqref{eqn:V} numerically,
we have calculated the components and
also the amplitude of the electric field $E = \sqrt{E_x^2 + E_z^2}$ over an interval of $x$ a few $|\delta|$ in length and $z$ varying from $0$ to $d$.
Our results for $E = E(x, z)$ 
for two doping levels, corresponding to $\lambda_p = 0$ (undoped SS) and $\lambda_p = a$ (doped SS)
are illustrated by the false color plots in Fig.~\ref{fig:gh}(c).
These plots are superimposed on perspective projections of the two slabs (doped and undoped),
which are placed next to each other for easy comparison.
The remaining parameters of the calculations are $\delta = -2.2 d$ and $a = 0.1 d$. 
We see that a finite shift of the ``hot stripes'' at the top surface $z = d$ exists in the doped case.
This seems to vindicate our intuition but
actually the situation is a bit more subtle.
The problem is that the momentum distribution of our source [Eq.~\eqref{eqn:V_launch}] is 
very different from what we assumed it to be in the beginning of our discussion of the GH effect.
This distribution
is not narrow and not centered at some finite $k^x$.
Instead, it has positive and negative $k^x$ harmonics of equal strength and
a long power-law tail at $|k^x| \gg 1 / a$.
The reason why the GH shift persists in our case is
the spatial separation of the $k^x$ harmonics:
due to the directionality of the HP$^2$ propagation,
the stripes to the left (right) of the launching points are created predominantly by negative (positive) $k_x$.
Since $\mathbf{l}$ has the same direction as $\mathbf{q} = (k^x, 0)$,
the stripes shift away from the origin on both sides of the $y$-axis.
A formal derivation of this result can be done by splitting the integral in Eq.~\eqref{eqn:V} into the $k_x > 0$ and the $k_x < 0$ parts and
identifying the relevant poles $k_x = q_n$ using contour integration methods.

From numerical experiments with different $a$, we found that
the largest shift of the stripes is obtained for $a \sim \lambda_p$.
This can be explained by arguing that the shift is maximized when the characteristic $k^x \sim \pi / a$ contributing to the integral in Eq.~\eqref{eqn:V} is close to the momentum $\pi / \lambda_p$
at which $l = l_{\mathrm{max}}$ in Eq.~\eqref{eqn:l}.

Experimental detection of the ``hot stripes'' and their doping-dependent GH shift is possible via the s-SNOM imaging.
This technique involves measuring the light scattered by the tip of an atomic force microscope brought to the sample and scanned along its surface.~\cite{Keilmann2004nfm, Atkin2012noi}
Using clever signal processing methods,
it is possible to isolate the genuine near-field component of this scattered light,
which originates from conversion of evanescent electromagnetic waves emanating from the sample into free-space photons.
In the proposed experiment, the evanescent waves are due to the HP$^2$ modes launched by the split gate.
The spatial resolution of the s-SNOM imaging is set by
the tip curvature radius $R$.
For typical $R = 20$--$40\unit{nm}$,
it is barely sufficient to observe the predicted GH shifts in hBN/G,
Fig.~\ref{fig:gh2}(b).
Nevertheless, detecting the cumulative shift after several stripe periods should be feasible.
The prior success of s-SNOM imaging experiments of surface plasmons and polaritons in graphene and hBN structures~\cite{Fei2011, Chen2012, Dai07032014, Dai2015sfg, Li2015hpp, Dai2015goh, Shi2015apr, Ni2015}
gives us a firm confidence in this approach.
Note that if a doped graphene layer only partially covers the top surface of hBN,
one literally gets the situation depicted in Fig.~\ref{fig:gh}(c),
where the doped and undoped regions are positioned side by side.

In the case of \BiSe* where the GH shift $\sim 200\unit{nm}$
[Fig.~\ref{fig:gh2}(a)] is much larger,
the spatial resolution of the s-SNOM is even less of an issue.
The main obstacle is the scant availability of suitable THz sources.
We are optimistic that in a near future this problem can be overcome as well.

\section{Summary and outlook}
\label{sec:conclusions}

Recent experiments~\cite{DiPietro2013, Autore2015ppi} have shown that coupling between Dirac plasmons and bulk phonons of bismuth-based TIs should be strong.
In this paper we have studied this interaction
taking into account the anisotropic phonon spectrum of such TIs.
We have predicted that a TI slab can act as a tunable waveguide for phonon polaritons, with the doping of the surface states being the tuning parameter.
In additional to the change in dispersion,
the phonon-plasmon coupling can cause measurable real-space shifts of the polariton rays. 
Similar phenomena have been recently studied in artificial structures made by stacking graphene layers on top of hBN.
The present work indicates
that the TIs are a promising alternative platform
for realizing highly tunable, strongly confined, low-loss electromagnetic modes in a \textit{natural} material.
Additionally, while hBN/G waveguides operate in mid-infrared frequencies,
\BiSe* and similar compounds extend the same functionality to
the technologically important THz domain.

We envision several directions for further work in this field.
One is to attempt a multi-source coherent control of polariton emission and propagation using ultrafast laser pulses.
A variety of such techniques has been developed~\cite{Feurer2007tp}
in the context of THz polaritonics of
LiNbO$_3$ and LiTiO$_3$.
(Incidentally, a theoretical proposal~\cite{Jin2013tpf} of integrating graphene into such materials would lead to polariton waveguides similar in functionality and perhaps also tunability
to those studied in the present work.)
Another intriguing direction is
to explore oscillating spin currents
which were predicted to accompany charge density currents produced by Dirac plasmons.~\cite{Raghu2010}
It may be also interesting to study the effect of optical hyperbolicity~\cite{Esslinger2014} on the high-energy bulk plasmons of the TIs.~\cite{Cha2013, Ou2014}
Finally, it may be worthwhile to investigate
new applications that can be enabled
by tunable hyperbolic polaritons.
Harnessing such types of modes for hyperlensing~\cite{Jacob2006, Salandrino2006ffs, Liu2007ffo} or focusing~\cite{Dai2015sfg, Li2015hpp} has been widely discussed.
The present work shows that the GH effect and its dependence
on doping and dielectric environment of the TI can be another avenue for applications,
for example, THz chemical sensing or characterization of spatially inhomogeneous TI samples.
We hope our work can stimulate these and other future studies.

\begin{acknowledgments}

This work is supported by
the University of California Office of the President and 
by the ONR.

\end{acknowledgments}

\appendix
\section{Near-field spectra}
\label{sec:SNOM}

A fully realistic modeling of the s-SNOM imaging experiments proposed in Sec.~\ref{sec:GH} is an unwieldy task requiring a repeated solution of the Maxwell equations for a system with complicated material properties, a hierarchy of widely different length scales, and no special symmetries.
In this Appendix we present some results of less ambitious calculations that simulate a simpler structure depicted in Fig.~\ref{fig:model_b}(b).
Although no split gate is present in this structure,
the measured signal is still expected to reveal characteristics of the collective modes.
In this case these modes are excited by the sharp tip itself.
Hence, the tip plays the role of both the launcher and the detector of the HP$^3$ modes.
Unfortunately, this implies that only the local response can be measured, which is a superposition of responses due to a distribution of momenta up to $q_t \sim 1 / R$ rather than one specific $q$.

\begin{figure}[b]
\includegraphics[width=7.5cm]{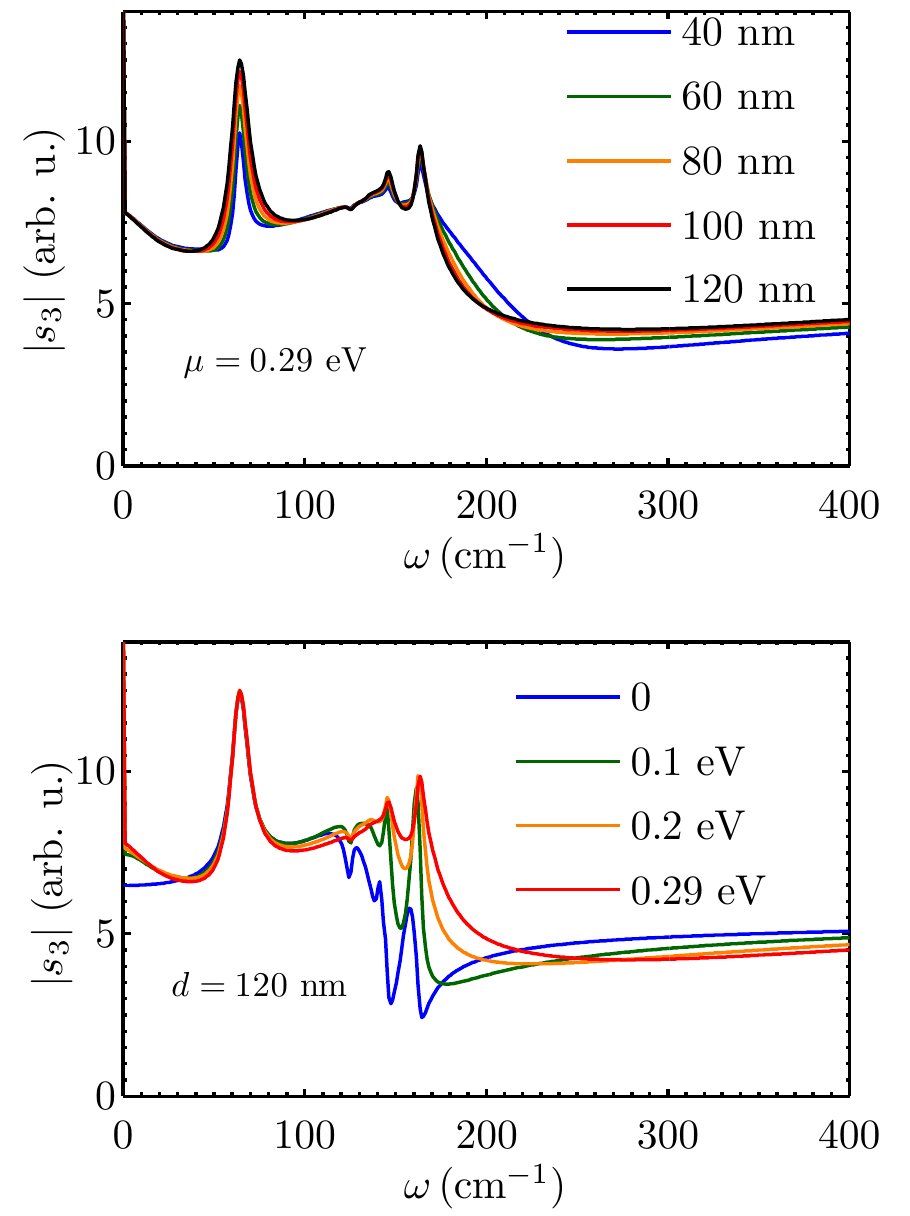}
\caption{(Color online) Simulation of the s-SNOM signal $s_3$ for \BiSe*
slabs on a substrate with $\epsilon_s = 10$.
(a) Fixed $\mu = 0.29\unit{eV}$ and different $d$.  
(b) Fixed $d = 120\unit{nm}$ and different $\mu$.
}
\label{fig:s3}
\end{figure}

We assume that the TI slab and the substrate are
infinite and uniform in $x$ and $y$ coordinates,
so that the imaging capability of the s-SNOM is irrelevant.
Instead, the quantity of interest is the
frequency dependence of the measured near-field signal $s(\omega)$.
A few more explanations about our calculational scheme are in order.
We model the tip as a metallic spheroid with the curvature radius $R = 40\unit{nm}$ and total length $720\unit{nm}$.
We use the quasistatic approximation but include the radiative corrections included perturbatively.
This model~\cite{Zhang2012, Jiang2015gsm} has been successful for simulating many recent s-SNOM experiments,
and should be especially suitable in the THz domain
where no antenna resonances or other strong retardation effects~\cite{McLeod2014mqt} should appear.
Our calculations incorporate the so-called far-field factors,~\cite{Zhang2012, Jiang2015gsm, McLeod2014mqt}
which are expressed in terms of $r_{P}(q, \omega)$ at $q \sim \omega / c$.
This factors
account for the fact that the incident wave is originally created by
a far-field source and the scattered wave is ultimately  
measured by a far-field detector.
Finally,
what we compute is not the full scattering amplitude $s$ but its third harmonic $s_3$,
which is what experimentalists typically report.
The idea is that in the experiment the tip is made to oscillate at some low frequency $\Omega$,
so that $s$ is periodic with this fundamental tapping frequency.
The third Fourier harmonic of $s$, which is $s_3$,
gives a good representation of the genuine near-field signal.

Naively, one can think of $s_3(\omega)$ as a weighted average
of the surface reflectivity $r_{P}(q, \omega)$
over $q$. The weighting function has a broad maximum near $q = q_t$,
which in this case is equal to
$q_t = 0.025\unit{nm}^{-1}$
[the dashed lines in Fig.~\ref{fig:rp_slab}].
The presence of strong maxima of $\Im\mathrm{m}\,r_{P}$
due to collective modes
with momenta $q \lesssim q_t$
tends to enhance $s_3(\omega)$.
In a more rigorous picture,~\cite{Jiang2015gsm}
the maxima of $s_3(\omega)$ correspond not to the resonances of the sample alone but to those of the coupled tip-sample system.
The coupling can decrease the resonance frequencies
by as much as~\cite{Zhang2012, Jiang2015gsm, McLeod2014mqt} $10$--$20\unit{cm}^{-1}$
compared to those seen in $\Im\mathrm{m}\,r_{P}$ maps.

Our results for \BiSe* slabs of various thickness $d$ and chemical potential $\mu$ are shown in Fig.~\ref{fig:s3}.
Pairs of distinct peaks as well as smaller additional features are readily seen.
In each trace,
the stronger and sharper peak is located
close to $\omega_{to, 1}^\bot = 64\unit{cm}^{-1}$.
The height of this peak decreases as $d$ decreases [Fig.~\ref{fig:s3}(a)].
However, its position is independent of $d$ [Fig.~\ref{fig:s3}(a)]
or $\mu$ [Fig.~\ref{fig:s3}(b)],
which suggests that it is not related to the dispersive HP$^3$ modes.
Indeed, we have verified that this prominent peak is almost entirely due to the far-field factor $|1 + r_P|^2$, which has a narrow maximum at $\omega_{to, 1}^\bot$ where
$r_{P} \approx 1$.

Each of the doped samples also produces smaller peaks in $s_3(\omega)$,
of which the most prominent ones are those located near $\omega = 146\unit{cm}^{-1}$ and 
$\omega = 163\unit{cm}^{-1}$,
the upper boundaries of regions B and C of Fig.~\ref{fig:epsi}.
The position and especially the strength of the peaks is $\mu$-dependent.
As $\mu$ increases,
the peaks grow in height and gradually shift to higher frequencies,
see Fig.~\ref{fig:s3}(b).
These peaks are due to the surface modes:
the $n = 1$ mode of region B and the $n = 0$ mode just above region C,
see Fig.~\ref{fig:rp_slab}(d, e).
The increase of the peak heights with $\mu$ can be qualitatively explained by the increase of the absolute value of $r_P$.
The shift in position is unfortunately more difficult to interpret without
a better understanding of the effective weighting function that relates $\Im\mathrm{m}\,r_{P}(q, \omega)$ to $s_3(\omega)$.

While $\mu > 0$ traces are due to combined action of plasmons and phonon-polaritons,
the $\mu = 0$ one is expected to reveal
the phonon-polariton response.
Interestingly, that trace exhibits a sharp dip at $\omega = 163\unit{cm}^{-1}$, see Fig.~\ref{fig:s3}(b).
We have checked that this dip is not caused by the far-field factor.
However, its relation to
the HP$^2$ modes of Fig.~\ref{fig:rp_slab}(d) is not obvious to us.

The thickness dependence of $s_3$ is illustrated in Fig.~\ref{fig:s3}(a).
As one can see, the near-field peak at $\omega = 163\unit{cm}^{-1}$
has a broad high-frequency side, which systematically expands as $d$ decreases.
This trend reflects the blue shift
of the $n = 0$ mode dispersion in thinner slabs, compare Fig.~\ref{fig:rp_slab}(e) and (f).

Overall, our simulations predict that the near-field response of \BiSe* slabs should exhibit systematic spectral changes with doping and thickness that are measurable by the s-SNOM.
Such experiments may provide insights into properties of tunable HP$^3$ modes of these novel systems.

\bibliography{collection}

\begin{thebibliography}{69}%
\makeatletter
\providecommand \@ifxundefined [1]{%
 \@ifx{#1\undefined}
}%
\providecommand \@ifnum [1]{%
 \ifnum #1\expandafter \@firstoftwo
 \else \expandafter \@secondoftwo
 \fi
}%
\providecommand \@ifx [1]{%
 \ifx #1\expandafter \@firstoftwo
 \else \expandafter \@secondoftwo
 \fi
}%
\providecommand \natexlab [1]{#1}%
\providecommand \enquote  [1]{``#1''}%
\providecommand \bibnamefont  [1]{#1}%
\providecommand \bibfnamefont [1]{#1}%
\providecommand \citenamefont [1]{#1}%
\providecommand \href@noop [0]{\@secondoftwo}%
\providecommand \href [0]{\begingroup \@sanitize@url \@href}%
\providecommand \@href[1]{\@@startlink{#1}\@@href}%
\providecommand \@@href[1]{\endgroup#1\@@endlink}%
\providecommand \@sanitize@url [0]{\catcode `\\12\catcode `\$12\catcode
  `\&12\catcode `\#12\catcode `\^12\catcode `\_12\catcode `\%12\relax}%
\providecommand \@@startlink[1]{}%
\providecommand \@@endlink[0]{}%
\providecommand \url  [0]{\begingroup\@sanitize@url \@url }%
\providecommand \@url [1]{\endgroup\@href {#1}{\urlprefix }}%
\providecommand \urlprefix  [0]{URL }%
\providecommand \Eprint [0]{\href }%
\providecommand \doibase [0]{http://dx.doi.org/}%
\providecommand \selectlanguage [0]{\@gobble}%
\providecommand \bibinfo  [0]{\@secondoftwo}%
\providecommand \bibfield  [0]{\@secondoftwo}%
\providecommand \translation [1]{[#1]}%
\providecommand \BibitemOpen [0]{}%
\providecommand \bibitemStop [0]{}%
\providecommand \bibitemNoStop [0]{.\EOS\space}%
\providecommand \EOS [0]{\spacefactor3000\relax}%
\providecommand \BibitemShut  [1]{\csname bibitem#1\endcsname}%
\let\auto@bib@innerbib\@empty
\bibitem [{\citenamefont {Hasan}\ and\ \citenamefont {Kane}(2010)}]{Hasan2010}%
  \BibitemOpen
  \bibfield  {author} {\bibinfo {author} {\bibfnamefont {M.~Z.}\ \bibnamefont
  {Hasan}}\ and\ \bibinfo {author} {\bibfnamefont {C.~L.}\ \bibnamefont
  {Kane}},\ }\href {\doibase 10.1103/RevModPhys.82.3045} {\bibfield  {journal}
  {\bibinfo  {journal} {{R}ev. {M}od. {P}hys.}\ }\textbf {\bibinfo {volume}
  {82}},\ \bibinfo {pages} {3045} (\bibinfo {year} {2010})}\BibitemShut
  {NoStop}%
\bibitem [{\citenamefont {Qi}\ and\ \citenamefont {Zhang}(2011)}]{Qi2011tia}%
  \BibitemOpen
  \bibfield  {author} {\bibinfo {author} {\bibfnamefont {X.-L.}\ \bibnamefont
  {Qi}}\ and\ \bibinfo {author} {\bibfnamefont {S.-C.}\ \bibnamefont {Zhang}},\
  }\href {\doibase 10.1103/RevModPhys.83.1057} {\bibfield  {journal} {\bibinfo
  {journal} {{R}ev. {M}od. {P}hys.}\ }\textbf {\bibinfo {volume} {83}},\
  \bibinfo {pages} {1057} (\bibinfo {year} {2011})}\BibitemShut {NoStop}%
\bibitem [{\citenamefont {Richter}\ \emph {et~al.}(1977)\citenamefont
  {Richter}, \citenamefont {K\"{o}hler},\ and\ \citenamefont
  {Becker}}]{Richter1977}%
  \BibitemOpen
  \bibfield  {author} {\bibinfo {author} {\bibfnamefont {W.}~\bibnamefont
  {Richter}}, \bibinfo {author} {\bibfnamefont {H.}~\bibnamefont {K\"{o}hler}},
  \ and\ \bibinfo {author} {\bibfnamefont {C.~R.}\ \bibnamefont {Becker}},\
  }\href {\doibase 10.1002/pssb.2220840226} {\bibfield  {journal} {\bibinfo
  {journal} {{P}hys. {S}tat. {S}ol. (b)}\ }\textbf {\bibinfo {volume} {84}},\
  \bibinfo {pages} {619} (\bibinfo {year} {1977})}\BibitemShut {NoStop}%
\bibitem [{\citenamefont {LaForge}\ \emph {et~al.}(2010)\citenamefont
  {LaForge}, \citenamefont {Frenzel}, \citenamefont {Pursley}, \citenamefont
  {Lin}, \citenamefont {Liu}, \citenamefont {Shi},\ and\ \citenamefont
  {Basov}}]{LaForge2010}%
  \BibitemOpen
  \bibfield  {author} {\bibinfo {author} {\bibfnamefont {A.~D.}\ \bibnamefont
  {LaForge}}, \bibinfo {author} {\bibfnamefont {A.}~\bibnamefont {Frenzel}},
  \bibinfo {author} {\bibfnamefont {B.~C.}\ \bibnamefont {Pursley}}, \bibinfo
  {author} {\bibfnamefont {T.}~\bibnamefont {Lin}}, \bibinfo {author}
  {\bibfnamefont {X.}~\bibnamefont {Liu}}, \bibinfo {author} {\bibfnamefont
  {J.}~\bibnamefont {Shi}}, \ and\ \bibinfo {author} {\bibfnamefont {D.~N.}\
  \bibnamefont {Basov}},\ }\href {\doibase 10.1103/PhysRevB.81.125120}
  {\bibfield  {journal} {\bibinfo  {journal} {{P}hys. {R}ev. {B}}\ }\textbf
  {\bibinfo {volume} {81}},\ \bibinfo {pages} {125120} (\bibinfo {year}
  {2010})}\BibitemShut {NoStop}%
\bibitem [{\citenamefont {Cheng}\ and\ \citenamefont {Ren}(2011)}]{Cheng2011}%
  \BibitemOpen
  \bibfield  {author} {\bibinfo {author} {\bibfnamefont {W.}~\bibnamefont
  {Cheng}}\ and\ \bibinfo {author} {\bibfnamefont {S.-F.}\ \bibnamefont
  {Ren}},\ }\href {\doibase 10.1103/PhysRevB.83.094301} {\bibfield  {journal}
  {\bibinfo  {journal} {{P}hys. {R}ev. {B}}\ }\textbf {\bibinfo {volume}
  {83}},\ \bibinfo {pages} {094301} (\bibinfo {year} {2011})}\BibitemShut
  {NoStop}%
\bibitem [{\citenamefont {Akrap}\ \emph {et~al.}(2012)\citenamefont {Akrap},
  \citenamefont {Tran}, \citenamefont {Ubaldini}, \citenamefont {Teyssier},
  \citenamefont {Giannini}, \citenamefont {van~der Marel}, \citenamefont
  {Lerch},\ and\ \citenamefont {Homes}}]{Akrap2012}%
  \BibitemOpen
  \bibfield  {author} {\bibinfo {author} {\bibfnamefont {A.}~\bibnamefont
  {Akrap}}, \bibinfo {author} {\bibfnamefont {M.}~\bibnamefont {Tran}},
  \bibinfo {author} {\bibfnamefont {A.}~\bibnamefont {Ubaldini}}, \bibinfo
  {author} {\bibfnamefont {J.}~\bibnamefont {Teyssier}}, \bibinfo {author}
  {\bibfnamefont {E.}~\bibnamefont {Giannini}}, \bibinfo {author}
  {\bibfnamefont {D.}~\bibnamefont {van~der Marel}}, \bibinfo {author}
  {\bibfnamefont {P.}~\bibnamefont {Lerch}}, \ and\ \bibinfo {author}
  {\bibfnamefont {C.~C.}\ \bibnamefont {Homes}},\ }\href {\doibase
  10.1103/PhysRevB.86.235207} {\bibfield  {journal} {\bibinfo  {journal}
  {{P}hys. {R}ev. {B}}\ }\textbf {\bibinfo {volume} {86}},\ \bibinfo {pages}
  {235207} (\bibinfo {year} {2012})}\BibitemShut {NoStop}%
\bibitem [{\citenamefont {Di~Pietro}\ \emph {et~al.}(2012)\citenamefont
  {Di~Pietro}, \citenamefont {Vitucci}, \citenamefont {Nicoletti},
  \citenamefont {Baldassarre}, \citenamefont {Calvani}, \citenamefont {Cava},
  \citenamefont {Hor}, \citenamefont {Schade},\ and\ \citenamefont
  {Lupi}}]{DiPietro2012ocb}%
  \BibitemOpen
  \bibfield  {author} {\bibinfo {author} {\bibfnamefont {P.}~\bibnamefont
  {Di~Pietro}}, \bibinfo {author} {\bibfnamefont {F.~M.}\ \bibnamefont
  {Vitucci}}, \bibinfo {author} {\bibfnamefont {D.}~\bibnamefont {Nicoletti}},
  \bibinfo {author} {\bibfnamefont {L.}~\bibnamefont {Baldassarre}}, \bibinfo
  {author} {\bibfnamefont {P.}~\bibnamefont {Calvani}}, \bibinfo {author}
  {\bibfnamefont {R.}~\bibnamefont {Cava}}, \bibinfo {author} {\bibfnamefont
  {Y.~S.}\ \bibnamefont {Hor}}, \bibinfo {author} {\bibfnamefont
  {U.}~\bibnamefont {Schade}}, \ and\ \bibinfo {author} {\bibfnamefont
  {S.}~\bibnamefont {Lupi}},\ }\href {\doibase 10.1103/PhysRevB.86.045439}
  {\bibfield  {journal} {\bibinfo  {journal} {{P}hys. {R}ev. {B}}\ }\textbf
  {\bibinfo {volume} {86}},\ \bibinfo {pages} {045439} (\bibinfo {year}
  {2012})}\BibitemShut {NoStop}%
\bibitem [{\citenamefont {{Di Pietro}}\ \emph {et~al.}(2013)\citenamefont {{Di
  Pietro}}, \citenamefont {Ortolani}, \citenamefont {Limaj}, \citenamefont {{Di
  Gaspare}}, \citenamefont {Giliberti}, \citenamefont {Giorgianni},
  \citenamefont {Brahlek}, \citenamefont {Bansal}, \citenamefont {Koirala},
  \citenamefont {Oh}, \citenamefont {Calvani},\ and\ \citenamefont
  {Lupi}}]{DiPietro2013}%
  \BibitemOpen
  \bibfield  {author} {\bibinfo {author} {\bibfnamefont {P.}~\bibnamefont {{Di
  Pietro}}}, \bibinfo {author} {\bibfnamefont {M.}~\bibnamefont {Ortolani}},
  \bibinfo {author} {\bibfnamefont {O.}~\bibnamefont {Limaj}}, \bibinfo
  {author} {\bibfnamefont {A.}~\bibnamefont {{Di Gaspare}}}, \bibinfo {author}
  {\bibfnamefont {V.}~\bibnamefont {Giliberti}}, \bibinfo {author}
  {\bibfnamefont {F.}~\bibnamefont {Giorgianni}}, \bibinfo {author}
  {\bibfnamefont {M.}~\bibnamefont {Brahlek}}, \bibinfo {author} {\bibfnamefont
  {N.}~\bibnamefont {Bansal}}, \bibinfo {author} {\bibfnamefont
  {N.}~\bibnamefont {Koirala}}, \bibinfo {author} {\bibfnamefont
  {S.}~\bibnamefont {Oh}}, \bibinfo {author} {\bibfnamefont {P.}~\bibnamefont
  {Calvani}}, \ and\ \bibinfo {author} {\bibfnamefont {S.}~\bibnamefont
  {Lupi}},\ }\href {\doibase 10.1038/nnano.2013.134} {\bibfield  {journal}
  {\bibinfo  {journal} {{N}at. {N}ano.}\ }\textbf {\bibinfo {volume} {8}},\
  \bibinfo {pages} {556} (\bibinfo {year} {2013})}\BibitemShut {NoStop}%
\bibitem [{\citenamefont {Wu}\ \emph {et~al.}(2013)\citenamefont {Wu},
  \citenamefont {Brahlek}, \citenamefont {Vald\'es~Aguilar}, \citenamefont
  {Stier}, \citenamefont {Morris}, \citenamefont {Lubashevsky}, \citenamefont
  {Bilbro}, \citenamefont {Bansal}, \citenamefont {Oh},\ and\ \citenamefont
  {Armitage}}]{Wu2013sct}%
  \BibitemOpen
  \bibfield  {author} {\bibinfo {author} {\bibfnamefont {L.}~\bibnamefont
  {Wu}}, \bibinfo {author} {\bibfnamefont {M.}~\bibnamefont {Brahlek}},
  \bibinfo {author} {\bibfnamefont {R.}~\bibnamefont {Vald\'es~Aguilar}},
  \bibinfo {author} {\bibfnamefont {A.~V.}\ \bibnamefont {Stier}}, \bibinfo
  {author} {\bibfnamefont {C.~M.}\ \bibnamefont {Morris}}, \bibinfo {author}
  {\bibfnamefont {Y.}~\bibnamefont {Lubashevsky}}, \bibinfo {author}
  {\bibfnamefont {L.~S.}\ \bibnamefont {Bilbro}}, \bibinfo {author}
  {\bibfnamefont {N.}~\bibnamefont {Bansal}}, \bibinfo {author} {\bibfnamefont
  {S.}~\bibnamefont {Oh}}, \ and\ \bibinfo {author} {\bibfnamefont {N.~P.}\
  \bibnamefont {Armitage}},\ }\href {\doibase 10.1038/nphys2647} {\bibfield
  {journal} {\bibinfo  {journal} {{N}at. {P}hys.}\ }\textbf {\bibinfo {volume}
  {9}},\ \bibinfo {pages} {410} (\bibinfo {year} {2013})}\BibitemShut {NoStop}%
\bibitem [{\citenamefont {Post}\ \emph {et~al.}(2013)\citenamefont {Post},
  \citenamefont {Chapler}, \citenamefont {He}, \citenamefont {Kou},
  \citenamefont {Wang},\ and\ \citenamefont {Basov}}]{Post2013tdb}%
  \BibitemOpen
  \bibfield  {author} {\bibinfo {author} {\bibfnamefont {K.~W.}\ \bibnamefont
  {Post}}, \bibinfo {author} {\bibfnamefont {B.~C.}\ \bibnamefont {Chapler}},
  \bibinfo {author} {\bibfnamefont {L.}~\bibnamefont {He}}, \bibinfo {author}
  {\bibfnamefont {X.}~\bibnamefont {Kou}}, \bibinfo {author} {\bibfnamefont
  {K.~L.}\ \bibnamefont {Wang}}, \ and\ \bibinfo {author} {\bibfnamefont
  {D.~N.}\ \bibnamefont {Basov}},\ }\href {\doibase 10.1103/PhysRevB.88.075121}
  {\bibfield  {journal} {\bibinfo  {journal} {{P}hys. {R}ev. {B}}\ }\textbf
  {\bibinfo {volume} {88}},\ \bibinfo {pages} {075121} (\bibinfo {year}
  {2013})}\BibitemShut {NoStop}%
\bibitem [{\citenamefont {Chapler}\ \emph {et~al.}(2014)\citenamefont
  {Chapler}, \citenamefont {Post}, \citenamefont {Richardella}, \citenamefont
  {Lee}, \citenamefont {Tao}, \citenamefont {Samarth},\ and\ \citenamefont
  {Basov}}]{Chapler2014ief}%
  \BibitemOpen
  \bibfield  {author} {\bibinfo {author} {\bibfnamefont {B.~C.}\ \bibnamefont
  {Chapler}}, \bibinfo {author} {\bibfnamefont {K.~W.}\ \bibnamefont {Post}},
  \bibinfo {author} {\bibfnamefont {A.~R.}\ \bibnamefont {Richardella}},
  \bibinfo {author} {\bibfnamefont {J.~S.}\ \bibnamefont {Lee}}, \bibinfo
  {author} {\bibfnamefont {J.}~\bibnamefont {Tao}}, \bibinfo {author}
  {\bibfnamefont {N.}~\bibnamefont {Samarth}}, \ and\ \bibinfo {author}
  {\bibfnamefont {D.~N.}\ \bibnamefont {Basov}},\ }\href {\doibase
  10.1103/PhysRevB.89.235308} {\bibfield  {journal} {\bibinfo  {journal}
  {{P}hys. {R}ev. {B}}\ }\textbf {\bibinfo {volume} {89}},\ \bibinfo {pages}
  {235308} (\bibinfo {year} {2014})}\BibitemShut {NoStop}%
\bibitem [{\citenamefont {Reijnders}\ \emph {et~al.}(2014)\citenamefont
  {Reijnders}, \citenamefont {Tian}, \citenamefont {Sandilands}, \citenamefont
  {Pohl}, \citenamefont {Kivlichan}, \citenamefont {Zhao}, \citenamefont {Jia},
  \citenamefont {Charles}, \citenamefont {Cava}, \citenamefont {Alidoust},
  \citenamefont {Xu}, \citenamefont {Neupane}, \citenamefont {Hasan},
  \citenamefont {Wang}, \citenamefont {Cheong},\ and\ \citenamefont
  {Burch}}]{Reijnders2014}%
  \BibitemOpen
  \bibfield  {author} {\bibinfo {author} {\bibfnamefont {A.~A.}\ \bibnamefont
  {Reijnders}}, \bibinfo {author} {\bibfnamefont {Y.}~\bibnamefont {Tian}},
  \bibinfo {author} {\bibfnamefont {L.~J.}\ \bibnamefont {Sandilands}},
  \bibinfo {author} {\bibfnamefont {G.}~\bibnamefont {Pohl}}, \bibinfo {author}
  {\bibfnamefont {I.~D.}\ \bibnamefont {Kivlichan}}, \bibinfo {author}
  {\bibfnamefont {S.~Y.~F.}\ \bibnamefont {Zhao}}, \bibinfo {author}
  {\bibfnamefont {S.}~\bibnamefont {Jia}}, \bibinfo {author} {\bibfnamefont
  {M.~E.}\ \bibnamefont {Charles}}, \bibinfo {author} {\bibfnamefont {R.~J.}\
  \bibnamefont {Cava}}, \bibinfo {author} {\bibfnamefont {N.}~\bibnamefont
  {Alidoust}}, \bibinfo {author} {\bibfnamefont {S.}~\bibnamefont {Xu}},
  \bibinfo {author} {\bibfnamefont {M.}~\bibnamefont {Neupane}}, \bibinfo
  {author} {\bibfnamefont {M.~Z.}\ \bibnamefont {Hasan}}, \bibinfo {author}
  {\bibfnamefont {X.}~\bibnamefont {Wang}}, \bibinfo {author} {\bibfnamefont
  {S.~W.}\ \bibnamefont {Cheong}}, \ and\ \bibinfo {author} {\bibfnamefont
  {K.~S.}\ \bibnamefont {Burch}},\ }\href {\doibase 10.1103/PhysRevB.89.075138}
  {\bibfield  {journal} {\bibinfo  {journal} {{P}hys. {R}ev. {B}}\ }\textbf
  {\bibinfo {volume} {89}},\ \bibinfo {pages} {075138} (\bibinfo {year}
  {2014})}\BibitemShut {NoStop}%
\bibitem [{\citenamefont {Autore}\ \emph
  {et~al.}(2015{\natexlab{a}})\citenamefont {Autore}, \citenamefont
  {Engelkamp}, \citenamefont {D'Apuzzo}, \citenamefont {Gaspare}, \citenamefont
  {Pietro}, \citenamefont {Vecchio}, \citenamefont {Brahlek}, \citenamefont
  {Koirala}, \citenamefont {Oh},\ and\ \citenamefont {Lupi}}]{Autore2015omb}%
  \BibitemOpen
  \bibfield  {author} {\bibinfo {author} {\bibfnamefont {M.}~\bibnamefont
  {Autore}}, \bibinfo {author} {\bibfnamefont {H.}~\bibnamefont {Engelkamp}},
  \bibinfo {author} {\bibfnamefont {F.}~\bibnamefont {D'Apuzzo}}, \bibinfo
  {author} {\bibfnamefont {A.~D.}\ \bibnamefont {Gaspare}}, \bibinfo {author}
  {\bibfnamefont {P.~D.}\ \bibnamefont {Pietro}}, \bibinfo {author}
  {\bibfnamefont {I.~L.}\ \bibnamefont {Vecchio}}, \bibinfo {author}
  {\bibfnamefont {M.}~\bibnamefont {Brahlek}}, \bibinfo {author} {\bibfnamefont
  {N.}~\bibnamefont {Koirala}}, \bibinfo {author} {\bibfnamefont
  {S.}~\bibnamefont {Oh}}, \ and\ \bibinfo {author} {\bibfnamefont
  {S.}~\bibnamefont {Lupi}},\ }\href {\doibase 10.1021/acsphotonics.5b00036}
  {\bibfield  {journal} {\bibinfo  {journal} {{ACS} {P}hoton.}\ }\textbf
  {\bibinfo {volume} {Article ASAP}} (\bibinfo {year} {2015}{\natexlab{a}}),\
  10.1021/acsphotonics.5b00036}\BibitemShut {NoStop}%
\bibitem [{\citenamefont {Autore}\ \emph
  {et~al.}(2015{\natexlab{b}})\citenamefont {Autore}, \citenamefont {D'Apuzzo},
  \citenamefont {Di~Gaspare}, \citenamefont {Giliberti}, \citenamefont {Limaj},
  \citenamefont {Roy}, \citenamefont {Brahlek}, \citenamefont {Koirala},
  \citenamefont {Oh}, \citenamefont {Garc\'{\i}a~de Abajo},\ and\ \citenamefont
  {Lupi}}]{Autore2015ppi}%
  \BibitemOpen
  \bibfield  {author} {\bibinfo {author} {\bibfnamefont {M.}~\bibnamefont
  {Autore}}, \bibinfo {author} {\bibfnamefont {F.}~\bibnamefont {D'Apuzzo}},
  \bibinfo {author} {\bibfnamefont {A.}~\bibnamefont {Di~Gaspare}}, \bibinfo
  {author} {\bibfnamefont {V.}~\bibnamefont {Giliberti}}, \bibinfo {author}
  {\bibfnamefont {O.}~\bibnamefont {Limaj}}, \bibinfo {author} {\bibfnamefont
  {P.}~\bibnamefont {Roy}}, \bibinfo {author} {\bibfnamefont {M.}~\bibnamefont
  {Brahlek}}, \bibinfo {author} {\bibfnamefont {N.}~\bibnamefont {Koirala}},
  \bibinfo {author} {\bibfnamefont {S.}~\bibnamefont {Oh}}, \bibinfo {author}
  {\bibfnamefont {F.~J.}\ \bibnamefont {Garc\'{\i}a~de Abajo}}, \ and\ \bibinfo
  {author} {\bibfnamefont {S.}~\bibnamefont {Lupi}},\ }\href {\doibase
  10.1002/adom.201400513} {\bibfield  {journal} {\bibinfo  {journal} {{A}dv.
  {O}pt. {M}at.}\ ,\ \bibinfo {pages} {XXXXX}} (\bibinfo {year}
  {2015}{\natexlab{b}})}\BibitemShut {NoStop}%
\bibitem [{\citenamefont {Post}\ \emph {et~al.}(2015)\citenamefont {Post},
  \citenamefont {Chapler}, \citenamefont {Liu}, \citenamefont {Wu},
  \citenamefont {Stinson}, \citenamefont {Goldflam}, \citenamefont
  {Richardella}, \citenamefont {Lee}, \citenamefont {Reijnders}, \citenamefont
  {Burch}, \citenamefont {Fogler}, \citenamefont {Samarth},\ and\ \citenamefont
  {Basov}}]{Post2015src}%
  \BibitemOpen
  \bibfield  {author} {\bibinfo {author} {\bibfnamefont {K.~W.}\ \bibnamefont
  {Post}}, \bibinfo {author} {\bibfnamefont {B.~C.}\ \bibnamefont {Chapler}},
  \bibinfo {author} {\bibfnamefont {M.~K.}\ \bibnamefont {Liu}}, \bibinfo
  {author} {\bibfnamefont {J.~S.}\ \bibnamefont {Wu}}, \bibinfo {author}
  {\bibfnamefont {H.~T.}\ \bibnamefont {Stinson}}, \bibinfo {author}
  {\bibfnamefont {M.~D.}\ \bibnamefont {Goldflam}}, \bibinfo {author}
  {\bibfnamefont {A.~R.}\ \bibnamefont {Richardella}}, \bibinfo {author}
  {\bibfnamefont {J.~S.}\ \bibnamefont {Lee}}, \bibinfo {author} {\bibfnamefont
  {A.~A.}\ \bibnamefont {Reijnders}}, \bibinfo {author} {\bibfnamefont {K.~S.}\
  \bibnamefont {Burch}}, \bibinfo {author} {\bibfnamefont {M.~M.}\ \bibnamefont
  {Fogler}}, \bibinfo {author} {\bibfnamefont {N.}~\bibnamefont {Samarth}}, \
  and\ \bibinfo {author} {\bibfnamefont {D.~N.}\ \bibnamefont {Basov}},\
  }\href@noop {} {\bibfield  {journal} {\bibinfo  {journal} {{P}hys. {R}ev.
  {L}ett.}\ }\textbf {\bibinfo {volume} {XX}},\ \bibinfo {pages} {XXXXX}
  (\bibinfo {year} {2015})}\BibitemShut {NoStop}%
\bibitem [{\citenamefont {Guo}\ \emph {et~al.}(2012)\citenamefont {Guo},
  \citenamefont {Newman},\ and\ \citenamefont {Jacob}}]{Guo2012}%
  \BibitemOpen
  \bibfield  {author} {\bibinfo {author} {\bibfnamefont {Y.}~\bibnamefont
  {Guo}}, \bibinfo {author} {\bibfnamefont {C.~L.}\ \bibnamefont {Newman},
  \bibfnamefont {W.~Cortes}}, \ and\ \bibinfo {author} {\bibfnamefont
  {Z.}~\bibnamefont {Jacob}},\ }\href {\doibase 10.1155/2012/452502} {\bibfield
   {journal} {\bibinfo  {journal} {{A}dv. {O}pto{E}lectron.}\ }\textbf
  {\bibinfo {volume} {2012}},\ \bibinfo {pages} {452502} (\bibinfo {year}
  {2012})}\BibitemShut {NoStop}%
\bibitem [{\citenamefont {Poddubny}\ \emph {et~al.}(2013)\citenamefont
  {Poddubny}, \citenamefont {Iorsh}, \citenamefont {Belov},\ and\ \citenamefont
  {Kivshar}}]{Poddubny2013}%
  \BibitemOpen
  \bibfield  {author} {\bibinfo {author} {\bibfnamefont {A.}~\bibnamefont
  {Poddubny}}, \bibinfo {author} {\bibfnamefont {I.}~\bibnamefont {Iorsh}},
  \bibinfo {author} {\bibfnamefont {P.}~\bibnamefont {Belov}}, \ and\ \bibinfo
  {author} {\bibfnamefont {Y.}~\bibnamefont {Kivshar}},\ }\href {\doibase
  10.1038/nphoton.2013.243} {\bibfield  {journal} {\bibinfo  {journal} {{N}at.
  {P}hoton.}\ }\textbf {\bibinfo {volume} {7}},\ \bibinfo {pages} {948}
  (\bibinfo {year} {2013})}\BibitemShut {NoStop}%
\bibitem [{\citenamefont {Sun}\ \emph {et~al.}(2014)\citenamefont {Sun},
  \citenamefont {Litchinitser},\ and\ \citenamefont {Zhou}}]{Sun2014}%
  \BibitemOpen
  \bibfield  {author} {\bibinfo {author} {\bibfnamefont {J.}~\bibnamefont
  {Sun}}, \bibinfo {author} {\bibfnamefont {N.~M.}\ \bibnamefont
  {Litchinitser}}, \ and\ \bibinfo {author} {\bibfnamefont {J.}~\bibnamefont
  {Zhou}},\ }\href {\doibase 10.1021/ph4000983} {\bibfield  {journal} {\bibinfo
   {journal} {{ACS} {P}hoton.}\ }\textbf {\bibinfo {volume} {1}},\ \bibinfo
  {pages} {293} (\bibinfo {year} {2014})}\BibitemShut {NoStop}%
\bibitem [{\citenamefont {Esslinger}\ \emph {et~al.}(2014)\citenamefont
  {Esslinger}, \citenamefont {Vogelgesang}, \citenamefont {Talebi},
  \citenamefont {Khunsin}, \citenamefont {Gehring}, \citenamefont {de~Zuani},
  \citenamefont {Gompf},\ and\ \citenamefont {Kern}}]{Esslinger2014}%
  \BibitemOpen
  \bibfield  {author} {\bibinfo {author} {\bibfnamefont {M.}~\bibnamefont
  {Esslinger}}, \bibinfo {author} {\bibfnamefont {R.}~\bibnamefont
  {Vogelgesang}}, \bibinfo {author} {\bibfnamefont {N.}~\bibnamefont {Talebi}},
  \bibinfo {author} {\bibfnamefont {W.}~\bibnamefont {Khunsin}}, \bibinfo
  {author} {\bibfnamefont {P.}~\bibnamefont {Gehring}}, \bibinfo {author}
  {\bibfnamefont {S.}~\bibnamefont {de~Zuani}}, \bibinfo {author}
  {\bibfnamefont {B.}~\bibnamefont {Gompf}}, \ and\ \bibinfo {author}
  {\bibfnamefont {K.}~\bibnamefont {Kern}},\ }\href {\doibase
  10.1021/ph500296e} {\bibfield  {journal} {\bibinfo  {journal} {{ACS}
  {P}hotonics}\ }\textbf {\bibinfo {volume} {1}},\ \bibinfo {pages} {1285}
  (\bibinfo {year} {2014})}\BibitemShut {NoStop}%
\bibitem [{\citenamefont {Dai}\ \emph {et~al.}(2014)\citenamefont {Dai},
  \citenamefont {Fei}, \citenamefont {Ma}, \citenamefont {Rodin}, \citenamefont
  {Wagner}, \citenamefont {McLeod}, \citenamefont {Liu}, \citenamefont
  {Gannett}, \citenamefont {Regan}, \citenamefont {Watanabe}, \citenamefont
  {Taniguchi}, \citenamefont {Thiemens}, \citenamefont {Dominguez},
  \citenamefont {Castro~Neto}, \citenamefont {Zettl}, \citenamefont {Keilmann},
  \citenamefont {Jarillo-Herrero}, \citenamefont {Fogler},\ and\ \citenamefont
  {Basov}}]{Dai07032014}%
  \BibitemOpen
  \bibfield  {author} {\bibinfo {author} {\bibfnamefont {S.}~\bibnamefont
  {Dai}}, \bibinfo {author} {\bibfnamefont {Z.}~\bibnamefont {Fei}}, \bibinfo
  {author} {\bibfnamefont {Q.}~\bibnamefont {Ma}}, \bibinfo {author}
  {\bibfnamefont {A.~S.}\ \bibnamefont {Rodin}}, \bibinfo {author}
  {\bibfnamefont {M.}~\bibnamefont {Wagner}}, \bibinfo {author} {\bibfnamefont
  {A.~S.}\ \bibnamefont {McLeod}}, \bibinfo {author} {\bibfnamefont {M.~K.}\
  \bibnamefont {Liu}}, \bibinfo {author} {\bibfnamefont {W.}~\bibnamefont
  {Gannett}}, \bibinfo {author} {\bibfnamefont {W.}~\bibnamefont {Regan}},
  \bibinfo {author} {\bibfnamefont {K.}~\bibnamefont {Watanabe}}, \bibinfo
  {author} {\bibfnamefont {T.}~\bibnamefont {Taniguchi}}, \bibinfo {author}
  {\bibfnamefont {M.}~\bibnamefont {Thiemens}}, \bibinfo {author}
  {\bibfnamefont {G.}~\bibnamefont {Dominguez}}, \bibinfo {author}
  {\bibfnamefont {A.~H.}\ \bibnamefont {Castro~Neto}}, \bibinfo {author}
  {\bibfnamefont {A.}~\bibnamefont {Zettl}}, \bibinfo {author} {\bibfnamefont
  {F.}~\bibnamefont {Keilmann}}, \bibinfo {author} {\bibfnamefont
  {P.}~\bibnamefont {Jarillo-Herrero}}, \bibinfo {author} {\bibfnamefont
  {M.~M.}\ \bibnamefont {Fogler}}, \ and\ \bibinfo {author} {\bibfnamefont
  {D.~N.}\ \bibnamefont {Basov}},\ }\href {\doibase 10.1126/science.1246833}
  {\bibfield  {journal} {\bibinfo  {journal} {{S}cience}\ }\textbf {\bibinfo
  {volume} {343}},\ \bibinfo {pages} {1125} (\bibinfo {year}
  {2014})}\BibitemShut {NoStop}%
\bibitem [{\citenamefont {Jacob}(2014)}]{Jacob2014npp}%
  \BibitemOpen
  \bibfield  {author} {\bibinfo {author} {\bibfnamefont {Z.}~\bibnamefont
  {Jacob}},\ }\href {\doibase 10.1038/nmat4149} {\bibfield  {journal} {\bibinfo
   {journal} {{N}at. {M}ater.}\ }\textbf {\bibinfo {volume} {13}},\ \bibinfo
  {pages} {1081} (\bibinfo {year} {2014})}\BibitemShut {NoStop}%
\bibitem [{\citenamefont {Caldwell}\ \emph {et~al.}(2014)\citenamefont
  {Caldwell}, \citenamefont {Kretinin}, \citenamefont {Chen}, \citenamefont
  {Giannini}, \citenamefont {Fogler}, \citenamefont {Francescato},
  \citenamefont {Ellis}, \citenamefont {Tischler}, \citenamefont {Woods},
  \citenamefont {Giles}, \citenamefont {Hong}, \citenamefont {Watanabe},
  \citenamefont {Taniguchi}, \citenamefont {Maier},\ and\ \citenamefont
  {Novoselov}}]{Caldwell2014}%
  \BibitemOpen
  \bibfield  {author} {\bibinfo {author} {\bibfnamefont {J.~D.}\ \bibnamefont
  {Caldwell}}, \bibinfo {author} {\bibfnamefont {A.~V.}\ \bibnamefont
  {Kretinin}}, \bibinfo {author} {\bibfnamefont {Y.}~\bibnamefont {Chen}},
  \bibinfo {author} {\bibfnamefont {V.}~\bibnamefont {Giannini}}, \bibinfo
  {author} {\bibfnamefont {M.~M.}\ \bibnamefont {Fogler}}, \bibinfo {author}
  {\bibfnamefont {Y.}~\bibnamefont {Francescato}}, \bibinfo {author}
  {\bibfnamefont {C.~T.}\ \bibnamefont {Ellis}}, \bibinfo {author}
  {\bibfnamefont {J.~G.}\ \bibnamefont {Tischler}}, \bibinfo {author}
  {\bibfnamefont {C.~R.}\ \bibnamefont {Woods}}, \bibinfo {author}
  {\bibfnamefont {A.~J.}\ \bibnamefont {Giles}}, \bibinfo {author}
  {\bibfnamefont {M.}~\bibnamefont {Hong}}, \bibinfo {author} {\bibfnamefont
  {K.}~\bibnamefont {Watanabe}}, \bibinfo {author} {\bibfnamefont
  {T.}~\bibnamefont {Taniguchi}}, \bibinfo {author} {\bibfnamefont {S.~A.}\
  \bibnamefont {Maier}}, \ and\ \bibinfo {author} {\bibfnamefont {K.~S.}\
  \bibnamefont {Novoselov}},\ }\href {\doibase 10.1038/ncomms6221} {\bibfield
  {journal} {\bibinfo  {journal} {{N}at. {C}omm.}\ }\textbf {\bibinfo {volume}
  {5}},\ \bibinfo {pages} {5221} (\bibinfo {year} {2014})}\BibitemShut
  {NoStop}%
\bibitem [{\citenamefont {Dai}\ \emph {et~al.}(2015{\natexlab{a}})\citenamefont
  {Dai}, \citenamefont {Ma}, \citenamefont {Andersen}, \citenamefont {Mcleod},
  \citenamefont {Fei}, \citenamefont {Liu}, \citenamefont {Wagner},
  \citenamefont {Watanabe}, \citenamefont {Taniguchi}, \citenamefont
  {Thiemens}, \citenamefont {Keilmann}, \citenamefont {Jarillo-Herrero},
  \citenamefont {Fogler},\ and\ \citenamefont {Basov}}]{Dai2015sfg}%
  \BibitemOpen
  \bibfield  {author} {\bibinfo {author} {\bibfnamefont {S.}~\bibnamefont
  {Dai}}, \bibinfo {author} {\bibfnamefont {Q.}~\bibnamefont {Ma}}, \bibinfo
  {author} {\bibfnamefont {T.}~\bibnamefont {Andersen}}, \bibinfo {author}
  {\bibfnamefont {A.~S.}\ \bibnamefont {Mcleod}}, \bibinfo {author}
  {\bibfnamefont {Z.}~\bibnamefont {Fei}}, \bibinfo {author} {\bibfnamefont
  {M.~K.}\ \bibnamefont {Liu}}, \bibinfo {author} {\bibfnamefont
  {M.}~\bibnamefont {Wagner}}, \bibinfo {author} {\bibfnamefont
  {K.}~\bibnamefont {Watanabe}}, \bibinfo {author} {\bibfnamefont
  {T.}~\bibnamefont {Taniguchi}}, \bibinfo {author} {\bibfnamefont
  {M.}~\bibnamefont {Thiemens}}, \bibinfo {author} {\bibfnamefont
  {F.}~\bibnamefont {Keilmann}}, \bibinfo {author} {\bibfnamefont
  {P.}~\bibnamefont {Jarillo-Herrero}}, \bibinfo {author} {\bibfnamefont
  {M.~M.}\ \bibnamefont {Fogler}}, \ and\ \bibinfo {author} {\bibfnamefont
  {D.~N.}\ \bibnamefont {Basov}},\ }\href {\doibase 10.1038/ncomms7963}
  {\bibfield  {journal} {\bibinfo  {journal} {{N}at. {C}ommun.}\ }\textbf
  {\bibinfo {volume} {6}},\ \bibinfo {pages} {6963} (\bibinfo {year}
  {2015}{\natexlab{a}})}\BibitemShut {NoStop}%
\bibitem [{\citenamefont {Li}\ \emph {et~al.}(2015)\citenamefont {Li},
  \citenamefont {Lewin}, \citenamefont {Kretinin}, \citenamefont {Caldwell},
  \citenamefont {Novoselov}, \citenamefont {Taniguchi}, \citenamefont
  {Watanabe}, \citenamefont {Gaussmann},\ and\ \citenamefont
  {Taubner}}]{Li2015hpp}%
  \BibitemOpen
  \bibfield  {author} {\bibinfo {author} {\bibfnamefont {P.}~\bibnamefont
  {Li}}, \bibinfo {author} {\bibfnamefont {M.}~\bibnamefont {Lewin}}, \bibinfo
  {author} {\bibfnamefont {A.~V.}\ \bibnamefont {Kretinin}}, \bibinfo {author}
  {\bibfnamefont {J.~D.}\ \bibnamefont {Caldwell}}, \bibinfo {author}
  {\bibfnamefont {K.~S.}\ \bibnamefont {Novoselov}}, \bibinfo {author}
  {\bibfnamefont {T.}~\bibnamefont {Taniguchi}}, \bibinfo {author}
  {\bibfnamefont {K.}~\bibnamefont {Watanabe}}, \bibinfo {author}
  {\bibfnamefont {F.}~\bibnamefont {Gaussmann}}, \ and\ \bibinfo {author}
  {\bibfnamefont {T.}~\bibnamefont {Taubner}},\ }\href {\doibase
  10.1038/ncomms8507} {\bibfield  {journal} {\bibinfo  {journal} {{N}at.
  {C}omm.}\ }\textbf {\bibinfo {volume} {6}},\ \bibinfo {pages} {7507}
  (\bibinfo {year} {2015})}\BibitemShut {NoStop}%
\bibitem [{\citenamefont {Shi}\ \emph {et~al.}(2015)\citenamefont {Shi},
  \citenamefont {Bechtel}, \citenamefont {Berweger}, \citenamefont {Sun},
  \citenamefont {Zeng}, \citenamefont {Jin}, \citenamefont {Chang},
  \citenamefont {Martin}, \citenamefont {Raschke},\ and\ \citenamefont
  {Wang}}]{Shi2015apr}%
  \BibitemOpen
  \bibfield  {author} {\bibinfo {author} {\bibfnamefont {Z.}~\bibnamefont
  {Shi}}, \bibinfo {author} {\bibfnamefont {H.~A.}\ \bibnamefont {Bechtel}},
  \bibinfo {author} {\bibfnamefont {S.}~\bibnamefont {Berweger}}, \bibinfo
  {author} {\bibfnamefont {Y.}~\bibnamefont {Sun}}, \bibinfo {author}
  {\bibfnamefont {B.}~\bibnamefont {Zeng}}, \bibinfo {author} {\bibfnamefont
  {C.}~\bibnamefont {Jin}}, \bibinfo {author} {\bibfnamefont {H.}~\bibnamefont
  {Chang}}, \bibinfo {author} {\bibfnamefont {M.~C.}\ \bibnamefont {Martin}},
  \bibinfo {author} {\bibfnamefont {M.~B.}\ \bibnamefont {Raschke}}, \ and\
  \bibinfo {author} {\bibfnamefont {F.}~\bibnamefont {Wang}},\ }\href {\doibase
  10.1021/acsphotonics.5b00007} {\bibfield  {journal} {\bibinfo  {journal}
  {{ACS} {P}hoton.}\ }\textbf {\bibinfo {volume} {2}},\ \bibinfo {pages} {790}
  (\bibinfo {year} {2015})}\BibitemShut {NoStop}%
\bibitem [{\citenamefont {Brar}\ \emph {et~al.}(2013)\citenamefont {Brar},
  \citenamefont {Jang}, \citenamefont {Sherrott}, \citenamefont {Lopez},\ and\
  \citenamefont {Atwater}}]{Brar2013hct}%
  \BibitemOpen
  \bibfield  {author} {\bibinfo {author} {\bibfnamefont {V.~W.}\ \bibnamefont
  {Brar}}, \bibinfo {author} {\bibfnamefont {M.~S.}\ \bibnamefont {Jang}},
  \bibinfo {author} {\bibfnamefont {M.}~\bibnamefont {Sherrott}}, \bibinfo
  {author} {\bibfnamefont {J.~J.}\ \bibnamefont {Lopez}}, \ and\ \bibinfo
  {author} {\bibfnamefont {H.~A.}\ \bibnamefont {Atwater}},\ }\href {\doibase
  10.1021/nl400601c} {\bibfield  {journal} {\bibinfo  {journal} {{N}ano
  {L}ett.}\ }\textbf {\bibinfo {volume} {13}},\ \bibinfo {pages} {2541}
  (\bibinfo {year} {2013})}\BibitemShut {NoStop}%
\bibitem [{\citenamefont {Dai}\ \emph {et~al.}(2015{\natexlab{b}})\citenamefont
  {Dai}, \citenamefont {Ma}, \citenamefont {Liu}, \citenamefont {Andersen},
  \citenamefont {Fei}, \citenamefont {Goldflam}, \citenamefont {Wagner},
  \citenamefont {Watanabe}, \citenamefont {Taniguchi}, \citenamefont
  {Thiemens}, \citenamefont {Keilmann}, \citenamefont {Janssen}, \citenamefont
  {Zhu}, \citenamefont {Jarillo-Herrero}, \citenamefont {Fogler},\ and\
  \citenamefont {Basov}}]{Dai2015goh}%
  \BibitemOpen
  \bibfield  {author} {\bibinfo {author} {\bibfnamefont {S.}~\bibnamefont
  {Dai}}, \bibinfo {author} {\bibfnamefont {Q.}~\bibnamefont {Ma}}, \bibinfo
  {author} {\bibfnamefont {M.~K.}\ \bibnamefont {Liu}}, \bibinfo {author}
  {\bibfnamefont {T.}~\bibnamefont {Andersen}}, \bibinfo {author}
  {\bibfnamefont {Z.}~\bibnamefont {Fei}}, \bibinfo {author} {\bibfnamefont
  {M.~D.}\ \bibnamefont {Goldflam}}, \bibinfo {author} {\bibfnamefont
  {M.}~\bibnamefont {Wagner}}, \bibinfo {author} {\bibfnamefont
  {K.}~\bibnamefont {Watanabe}}, \bibinfo {author} {\bibfnamefont
  {T.}~\bibnamefont {Taniguchi}}, \bibinfo {author} {\bibfnamefont
  {M.}~\bibnamefont {Thiemens}}, \bibinfo {author} {\bibfnamefont
  {F.}~\bibnamefont {Keilmann}}, \bibinfo {author} {\bibfnamefont {G.~C.
  A.~M.}\ \bibnamefont {Janssen}}, \bibinfo {author} {\bibfnamefont {S.-E.}\
  \bibnamefont {Zhu}}, \bibinfo {author} {\bibfnamefont {P.}~\bibnamefont
  {Jarillo-Herrero}}, \bibinfo {author} {\bibfnamefont {M.~M.}\ \bibnamefont
  {Fogler}}, \ and\ \bibinfo {author} {\bibfnamefont {D.~N.}\ \bibnamefont
  {Basov}},\ }\href {\doibase 10.1038/nnano.2015.131} {\bibfield  {journal}
  {\bibinfo  {journal} {{N}at. {N}ano}\ }\textbf {\bibinfo {volume} {10}},\
  \bibinfo {pages} {682} (\bibinfo {year} {2015}{\natexlab{b}})}\BibitemShut
  {NoStop}%
\bibitem [{\citenamefont {Ni}\ \emph {et~al.}(2015)\citenamefont {Ni},
  \citenamefont {Wang}, \citenamefont {Wu}, \citenamefont {Fei}, \citenamefont
  {Goldflam}, \citenamefont {Keilmann}, \citenamefont {\"Ozyilmaz},
  \citenamefont {Castro~Neto}, \citenamefont {Xie}, \citenamefont {Fogler},\
  and\ \citenamefont {Basov}}]{Ni2015}%
  \BibitemOpen
  \bibfield  {author} {\bibinfo {author} {\bibfnamefont {G.~X.}\ \bibnamefont
  {Ni}}, \bibinfo {author} {\bibfnamefont {H.}~\bibnamefont {Wang}}, \bibinfo
  {author} {\bibfnamefont {J.~S.}\ \bibnamefont {Wu}}, \bibinfo {author}
  {\bibfnamefont {Z.}~\bibnamefont {Fei}}, \bibinfo {author} {\bibfnamefont
  {M.~D.}\ \bibnamefont {Goldflam}}, \bibinfo {author} {\bibfnamefont
  {F.}~\bibnamefont {Keilmann}}, \bibinfo {author} {\bibfnamefont
  {B.}~\bibnamefont {\"Ozyilmaz}}, \bibinfo {author} {\bibfnamefont {A.~H.}\
  \bibnamefont {Castro~Neto}}, \bibinfo {author} {\bibfnamefont {X.~M.}\
  \bibnamefont {Xie}}, \bibinfo {author} {\bibfnamefont {M.~M.}\ \bibnamefont
  {Fogler}}, \ and\ \bibinfo {author} {\bibfnamefont {D.~N.}\ \bibnamefont
  {Basov}},\ }\href@noop {} {\bibfield  {journal} {\bibinfo  {journal} {{N}at.
  {M}at.}\ }\textbf {\bibinfo {volume} {XXX}},\ \bibinfo {pages} {XXXXX}
  (\bibinfo {year} {2015})}\BibitemShut {NoStop}%
\bibitem [{\citenamefont {Tomadin}\ \emph {et~al.}(2014)\citenamefont
  {Tomadin}, \citenamefont {Guinea},\ and\ \citenamefont
  {Polini}}]{Tomadin2014gmp}%
  \BibitemOpen
  \bibfield  {author} {\bibinfo {author} {\bibfnamefont {A.}~\bibnamefont
  {Tomadin}}, \bibinfo {author} {\bibfnamefont {F.}~\bibnamefont {Guinea}}, \
  and\ \bibinfo {author} {\bibfnamefont {M.}~\bibnamefont {Polini}},\ }\href
  {\doibase 10.1103/PhysRevB.90.161406} {\bibfield  {journal} {\bibinfo
  {journal} {{P}hys. {R}ev. {B}}\ }\textbf {\bibinfo {volume} {90}},\ \bibinfo
  {pages} {161406} (\bibinfo {year} {2014})}\BibitemShut {NoStop}%
\bibitem [{\citenamefont {Castro~Neto}\ \emph {et~al.}(2009)\citenamefont
  {Castro~Neto}, \citenamefont {Guinea}, \citenamefont {Peres}, \citenamefont
  {Novoselov},\ and\ \citenamefont {Geim}}]{Castro2009}%
  \BibitemOpen
  \bibfield  {author} {\bibinfo {author} {\bibfnamefont {A.~H.}\ \bibnamefont
  {Castro~Neto}}, \bibinfo {author} {\bibfnamefont {F.}~\bibnamefont {Guinea}},
  \bibinfo {author} {\bibfnamefont {N.~M.~R.}\ \bibnamefont {Peres}}, \bibinfo
  {author} {\bibfnamefont {K.~S.}\ \bibnamefont {Novoselov}}, \ and\ \bibinfo
  {author} {\bibfnamefont {A.~K.}\ \bibnamefont {Geim}},\ }\href {\doibase
  10.1103/RevModPhys.81.109} {\bibfield  {journal} {\bibinfo  {journal} {{R}ev.
  {M}od. {P}hys.}\ }\textbf {\bibinfo {volume} {81}},\ \bibinfo {pages} {109}
  (\bibinfo {year} {2009})}\BibitemShut {NoStop}%
\bibitem [{\citenamefont {Hwang}\ and\ \citenamefont
  {Das~Sarma}(2009)}]{Hwang2009pms}%
  \BibitemOpen
  \bibfield  {author} {\bibinfo {author} {\bibfnamefont {E.~H.}\ \bibnamefont
  {Hwang}}\ and\ \bibinfo {author} {\bibfnamefont {S.}~\bibnamefont
  {Das~Sarma}},\ }\href {\doibase 10.1103/PhysRevB.80.205405} {\bibfield
  {journal} {\bibinfo  {journal} {{P}hys. {R}ev. {B}}\ }\textbf {\bibinfo
  {volume} {80}},\ \bibinfo {pages} {205405} (\bibinfo {year}
  {2009})}\BibitemShut {NoStop}%
\bibitem [{\citenamefont {Raghu}\ \emph {et~al.}(2010)\citenamefont {Raghu},
  \citenamefont {Chung}, \citenamefont {Qi},\ and\ \citenamefont
  {Zhang}}]{Raghu2010}%
  \BibitemOpen
  \bibfield  {author} {\bibinfo {author} {\bibfnamefont {S.}~\bibnamefont
  {Raghu}}, \bibinfo {author} {\bibfnamefont {S.~B.}\ \bibnamefont {Chung}},
  \bibinfo {author} {\bibfnamefont {X.-L.}\ \bibnamefont {Qi}}, \ and\ \bibinfo
  {author} {\bibfnamefont {S.-C.}\ \bibnamefont {Zhang}},\ }\href {\doibase
  10.1103/PhysRevLett.104.116401} {\bibfield  {journal} {\bibinfo  {journal}
  {{P}hys. {R}ev. {L}ett.}\ }\textbf {\bibinfo {volume} {104}},\ \bibinfo
  {pages} {116401} (\bibinfo {year} {2010})}\BibitemShut {NoStop}%
\bibitem [{\citenamefont {Fei}\ \emph {et~al.}(2011)\citenamefont {Fei},
  \citenamefont {Andreev}, \citenamefont {Bao}, \citenamefont {Zhang},
  \citenamefont {McLeod}, \citenamefont {Wang}, \citenamefont {Stewart},
  \citenamefont {Zhao}, \citenamefont {Dominguez}, \citenamefont {Thiemens},
  \citenamefont {Fogler}, \citenamefont {Tauber}, \citenamefont {Castro-Neto},
  \citenamefont {Lau}, \citenamefont {Keilmann},\ and\ \citenamefont
  {Basov}}]{Fei2011}%
  \BibitemOpen
  \bibfield  {author} {\bibinfo {author} {\bibfnamefont {Z.}~\bibnamefont
  {Fei}}, \bibinfo {author} {\bibfnamefont {G.~O.}\ \bibnamefont {Andreev}},
  \bibinfo {author} {\bibfnamefont {W.}~\bibnamefont {Bao}}, \bibinfo {author}
  {\bibfnamefont {L.~M.}\ \bibnamefont {Zhang}}, \bibinfo {author}
  {\bibfnamefont {A.~S.}\ \bibnamefont {McLeod}}, \bibinfo {author}
  {\bibfnamefont {C.}~\bibnamefont {Wang}}, \bibinfo {author} {\bibfnamefont
  {M.~K.}\ \bibnamefont {Stewart}}, \bibinfo {author} {\bibfnamefont
  {Z.}~\bibnamefont {Zhao}}, \bibinfo {author} {\bibfnamefont {G.}~\bibnamefont
  {Dominguez}}, \bibinfo {author} {\bibfnamefont {M.}~\bibnamefont {Thiemens}},
  \bibinfo {author} {\bibfnamefont {M.~M.}\ \bibnamefont {Fogler}}, \bibinfo
  {author} {\bibfnamefont {M.~J.}\ \bibnamefont {Tauber}}, \bibinfo {author}
  {\bibfnamefont {A.~H.}\ \bibnamefont {Castro-Neto}}, \bibinfo {author}
  {\bibfnamefont {C.~N.}\ \bibnamefont {Lau}}, \bibinfo {author} {\bibfnamefont
  {F.}~\bibnamefont {Keilmann}}, \ and\ \bibinfo {author} {\bibfnamefont
  {D.~N.}\ \bibnamefont {Basov}},\ }\href {\doibase 10.1021/nl202362d}
  {\bibfield  {journal} {\bibinfo  {journal} {{N}ano {L}ett.}\ }\textbf
  {\bibinfo {volume} {11}},\ \bibinfo {pages} {4701} (\bibinfo {year}
  {2011})}\BibitemShut {NoStop}%
\bibitem [{\citenamefont {Fei}\ \emph {et~al.}(2012)\citenamefont {Fei},
  \citenamefont {Rodin}, \citenamefont {Andreev}, \citenamefont {Bao},
  \citenamefont {McLeod}, \citenamefont {Wagner}, \citenamefont {Zhang},
  \citenamefont {Zhao}, \citenamefont {Thiemens}, \citenamefont {Dominguez},
  \citenamefont {Fogler}, \citenamefont {{Castro Neto}}, \citenamefont {Lau},
  \citenamefont {Keilmann},\ and\ \citenamefont {Basov}}]{Fei2012}%
  \BibitemOpen
  \bibfield  {author} {\bibinfo {author} {\bibfnamefont {Z.}~\bibnamefont
  {Fei}}, \bibinfo {author} {\bibfnamefont {A.~S.}\ \bibnamefont {Rodin}},
  \bibinfo {author} {\bibfnamefont {G.~O.}\ \bibnamefont {Andreev}}, \bibinfo
  {author} {\bibfnamefont {W.}~\bibnamefont {Bao}}, \bibinfo {author}
  {\bibfnamefont {A.~S.}\ \bibnamefont {McLeod}}, \bibinfo {author}
  {\bibfnamefont {M.}~\bibnamefont {Wagner}}, \bibinfo {author} {\bibfnamefont
  {L.~M.}\ \bibnamefont {Zhang}}, \bibinfo {author} {\bibfnamefont
  {Z.}~\bibnamefont {Zhao}}, \bibinfo {author} {\bibfnamefont {M.}~\bibnamefont
  {Thiemens}}, \bibinfo {author} {\bibfnamefont {G.}~\bibnamefont {Dominguez}},
  \bibinfo {author} {\bibfnamefont {M.~M.}\ \bibnamefont {Fogler}}, \bibinfo
  {author} {\bibfnamefont {A.~H.}\ \bibnamefont {{Castro Neto}}}, \bibinfo
  {author} {\bibfnamefont {C.~N.}\ \bibnamefont {Lau}}, \bibinfo {author}
  {\bibfnamefont {F.}~\bibnamefont {Keilmann}}, \ and\ \bibinfo {author}
  {\bibfnamefont {D.~N.}\ \bibnamefont {Basov}},\ }\href {\doibase
  10.1038/nature11253} {\bibfield  {journal} {\bibinfo  {journal} {{N}ature}\
  }\textbf {\bibinfo {volume} {487}},\ \bibinfo {pages} {82} (\bibinfo {year}
  {2012})}\BibitemShut {NoStop}%
\bibitem [{\citenamefont {Chen}\ \emph {et~al.}(2012)\citenamefont {Chen},
  \citenamefont {Badioli}, \citenamefont {Alonso-Gonz\'{a}lez}, \citenamefont
  {Thongrattanasiri}, \citenamefont {Huth}, \citenamefont {Osmond},
  \citenamefont {Spasenovi\'{c}}, \citenamefont {Centeno}, \citenamefont
  {Pesquera}, \citenamefont {Godignon}, \citenamefont {Elorza}, \citenamefont
  {Camara}, \citenamefont {{Garc\'{\i}a de Abajo}}, \citenamefont
  {Hillenbrand},\ and\ \citenamefont {Koppens}}]{Chen2012}%
  \BibitemOpen
  \bibfield  {author} {\bibinfo {author} {\bibfnamefont {J.}~\bibnamefont
  {Chen}}, \bibinfo {author} {\bibfnamefont {M.}~\bibnamefont {Badioli}},
  \bibinfo {author} {\bibfnamefont {P.}~\bibnamefont {Alonso-Gonz\'{a}lez}},
  \bibinfo {author} {\bibfnamefont {S.}~\bibnamefont {Thongrattanasiri}},
  \bibinfo {author} {\bibfnamefont {F.}~\bibnamefont {Huth}}, \bibinfo {author}
  {\bibfnamefont {J.}~\bibnamefont {Osmond}}, \bibinfo {author} {\bibfnamefont
  {M.}~\bibnamefont {Spasenovi\'{c}}}, \bibinfo {author} {\bibfnamefont
  {A.}~\bibnamefont {Centeno}}, \bibinfo {author} {\bibfnamefont
  {A.}~\bibnamefont {Pesquera}}, \bibinfo {author} {\bibfnamefont
  {P.}~\bibnamefont {Godignon}}, \bibinfo {author} {\bibfnamefont {A.~Z.}\
  \bibnamefont {Elorza}}, \bibinfo {author} {\bibfnamefont {N.}~\bibnamefont
  {Camara}}, \bibinfo {author} {\bibfnamefont {F.~J.}\ \bibnamefont
  {{Garc\'{\i}a de Abajo}}}, \bibinfo {author} {\bibfnamefont {R.}~\bibnamefont
  {Hillenbrand}}, \ and\ \bibinfo {author} {\bibfnamefont {F.~H.~L.}\
  \bibnamefont {Koppens}},\ }\href {\doibase 10.1038/nature11254} {\bibfield
  {journal} {\bibinfo  {journal} {{N}ature}\ }\textbf {\bibinfo {volume}
  {487}},\ \bibinfo {pages} {77} (\bibinfo {year} {2012})}\BibitemShut
  {NoStop}%
\bibitem [{\citenamefont {Grigorenko}\ \emph {et~al.}(2012)\citenamefont
  {Grigorenko}, \citenamefont {Polini},\ and\ \citenamefont
  {Novoselov}}]{Grigorenko2012gp}%
  \BibitemOpen
  \bibfield  {author} {\bibinfo {author} {\bibfnamefont {A.~N.}\ \bibnamefont
  {Grigorenko}}, \bibinfo {author} {\bibfnamefont {M.}~\bibnamefont {Polini}},
  \ and\ \bibinfo {author} {\bibfnamefont {K.~S.}\ \bibnamefont {Novoselov}},\
  }\href {\doibase 10.1038/nphoton.2012.262} {\bibfield  {journal} {\bibinfo
  {journal} {{N}ature {P}hoton.}\ }\textbf {\bibinfo {volume} {6}},\ \bibinfo
  {pages} {749} (\bibinfo {year} {2012})}\BibitemShut {NoStop}%
\bibitem [{\citenamefont {Profumo}\ \emph {et~al.}(2012)\citenamefont
  {Profumo}, \citenamefont {Asgari}, \citenamefont {Polini},\ and\
  \citenamefont {MacDonald}}]{Profumo2012}%
  \BibitemOpen
  \bibfield  {author} {\bibinfo {author} {\bibfnamefont {R.~E.~V.}\
  \bibnamefont {Profumo}}, \bibinfo {author} {\bibfnamefont {R.}~\bibnamefont
  {Asgari}}, \bibinfo {author} {\bibfnamefont {M.}~\bibnamefont {Polini}}, \
  and\ \bibinfo {author} {\bibfnamefont {A.~H.}\ \bibnamefont {MacDonald}},\
  }\href {\doibase 10.1103/PhysRevB.85.085443} {\bibfield  {journal} {\bibinfo
  {journal} {{P}hys. {R}ev. {B}}\ }\textbf {\bibinfo {volume} {85}},\ \bibinfo
  {pages} {085443} (\bibinfo {year} {2012})}\BibitemShut {NoStop}%
\bibitem [{\citenamefont {Garc\'{\i}a~de Abajo}(2014)}]{GarciadeAbajo2014gpc}%
  \BibitemOpen
  \bibfield  {author} {\bibinfo {author} {\bibfnamefont {F.~J.}\ \bibnamefont
  {Garc\'{\i}a~de Abajo}},\ }\href {\doibase 10.1021/ph400147y} {\bibfield
  {journal} {\bibinfo  {journal} {{ACS} {P}hoton.}\ }\textbf {\bibinfo {volume}
  {1}},\ \bibinfo {pages} {135} (\bibinfo {year} {2014})}\BibitemShut {NoStop}%
\bibitem [{\citenamefont {Basov}\ \emph {et~al.}(2014)\citenamefont {Basov},
  \citenamefont {Fogler}, \citenamefont {Lanzara}, \citenamefont {Wang},\ and\
  \citenamefont {Zhang}}]{Basov2014}%
  \BibitemOpen
  \bibfield  {author} {\bibinfo {author} {\bibfnamefont {D.~N.}\ \bibnamefont
  {Basov}}, \bibinfo {author} {\bibfnamefont {M.~M.}\ \bibnamefont {Fogler}},
  \bibinfo {author} {\bibfnamefont {A.}~\bibnamefont {Lanzara}}, \bibinfo
  {author} {\bibfnamefont {F.}~\bibnamefont {Wang}}, \ and\ \bibinfo {author}
  {\bibfnamefont {Y.}~\bibnamefont {Zhang}},\ }\href {\doibase
  10.1103/RevModPhys.86.959} {\bibfield  {journal} {\bibinfo  {journal} {{R}ev.
  {M}od. {P}hys.}\ }\textbf {\bibinfo {volume} {86}},\ \bibinfo {pages} {959}
  (\bibinfo {year} {2014})}\BibitemShut {NoStop}%
\bibitem [{\citenamefont {Stauber}\ \emph {et~al.}(2013)\citenamefont
  {Stauber}, \citenamefont {G\'omez-Santos},\ and\ \citenamefont
  {Brey}}]{Stauber2013}%
  \BibitemOpen
  \bibfield  {author} {\bibinfo {author} {\bibfnamefont {T.}~\bibnamefont
  {Stauber}}, \bibinfo {author} {\bibfnamefont {G.}~\bibnamefont
  {G\'omez-Santos}}, \ and\ \bibinfo {author} {\bibfnamefont {L.}~\bibnamefont
  {Brey}},\ }\href {\doibase 10.1103/PhysRevB.88.205427} {\bibfield  {journal}
  {\bibinfo  {journal} {{P}hys. {R}ev. {B}}\ }\textbf {\bibinfo {volume}
  {88}},\ \bibinfo {pages} {205427} (\bibinfo {year} {2013})}\BibitemShut
  {NoStop}%
\bibitem [{\citenamefont {Sch\"utky}\ \emph {et~al.}(2013)\citenamefont
  {Sch\"utky}, \citenamefont {Ertler}, \citenamefont {Tr\"ugler},\ and\
  \citenamefont {Hohenester}}]{Schutky2013spd}%
  \BibitemOpen
  \bibfield  {author} {\bibinfo {author} {\bibfnamefont {R.}~\bibnamefont
  {Sch\"utky}}, \bibinfo {author} {\bibfnamefont {C.}~\bibnamefont {Ertler}},
  \bibinfo {author} {\bibfnamefont {A.}~\bibnamefont {Tr\"ugler}}, \ and\
  \bibinfo {author} {\bibfnamefont {U.}~\bibnamefont {Hohenester}},\ }\href
  {\doibase 10.1103/PhysRevB.88.195311} {\bibfield  {journal} {\bibinfo
  {journal} {{P}hys. {R}ev. {B}}\ }\textbf {\bibinfo {volume} {88}},\ \bibinfo
  {pages} {195311} (\bibinfo {year} {2013})}\BibitemShut {NoStop}%
\bibitem [{\citenamefont {Qi}\ \emph {et~al.}(2014)\citenamefont {Qi},
  \citenamefont {Liu},\ and\ \citenamefont {Xie}}]{Qi2014spp}%
  \BibitemOpen
  \bibfield  {author} {\bibinfo {author} {\bibfnamefont {J.}~\bibnamefont
  {Qi}}, \bibinfo {author} {\bibfnamefont {H.}~\bibnamefont {Liu}}, \ and\
  \bibinfo {author} {\bibfnamefont {X.~C.}\ \bibnamefont {Xie}},\ }\href
  {\doibase 10.1103/PhysRevB.89.155420} {\bibfield  {journal} {\bibinfo
  {journal} {{P}hys. {R}ev. {B}}\ }\textbf {\bibinfo {volume} {89}},\ \bibinfo
  {pages} {155420} (\bibinfo {year} {2014})}\BibitemShut {NoStop}%
\bibitem [{\citenamefont {Li}\ \emph {et~al.}(2014)\citenamefont {Li},
  \citenamefont {Dai}, \citenamefont {Cui}, \citenamefont {Wang}, \citenamefont
  {Katmis}, \citenamefont {Wang}, \citenamefont {Le}, \citenamefont {Wu},\ and\
  \citenamefont {Zhu}}]{Li2014tts}%
  \BibitemOpen
  \bibfield  {author} {\bibinfo {author} {\bibfnamefont {M.}~\bibnamefont
  {Li}}, \bibinfo {author} {\bibfnamefont {Z.}~\bibnamefont {Dai}}, \bibinfo
  {author} {\bibfnamefont {W.}~\bibnamefont {Cui}}, \bibinfo {author}
  {\bibfnamefont {Z.}~\bibnamefont {Wang}}, \bibinfo {author} {\bibfnamefont
  {F.}~\bibnamefont {Katmis}}, \bibinfo {author} {\bibfnamefont
  {J.}~\bibnamefont {Wang}}, \bibinfo {author} {\bibfnamefont {P.}~\bibnamefont
  {Le}}, \bibinfo {author} {\bibfnamefont {L.}~\bibnamefont {Wu}}, \ and\
  \bibinfo {author} {\bibfnamefont {Y.}~\bibnamefont {Zhu}},\ }\href {\doibase
  10.1103/PhysRevB.89.235432} {\bibfield  {journal} {\bibinfo  {journal}
  {{P}hys. {R}ev. {B}}\ }\textbf {\bibinfo {volume} {89}},\ \bibinfo {pages}
  {235432} (\bibinfo {year} {2014})}\BibitemShut {NoStop}%
\bibitem [{\citenamefont {Stauber}(2014)}]{Stauber2014pds}%
  \BibitemOpen
  \bibfield  {author} {\bibinfo {author} {\bibfnamefont {T.}~\bibnamefont
  {Stauber}},\ }\href {\doibase 10.1088/0953-8984/26/12/123201} {\bibfield
  {journal} {\bibinfo  {journal} {{J}. {P}hys.: {C}ondens. {M}atter}\ }\textbf
  {\bibinfo {volume} {26}},\ \bibinfo {pages} {123201} (\bibinfo {year}
  {2014})}\BibitemShut {NoStop}%
\bibitem [{\citenamefont {Goos}\ and\ \citenamefont
  {H\"anchen}(1947)}]{Goos1947}%
  \BibitemOpen
  \bibfield  {author} {\bibinfo {author} {\bibfnamefont {F.}~\bibnamefont
  {Goos}}\ and\ \bibinfo {author} {\bibfnamefont {H.}~\bibnamefont
  {H\"anchen}},\ }\href {\doibase 10.1002/andp.19474360704} {\bibfield
  {journal} {\bibinfo  {journal} {{A}nn. {P}hys.}\ }\textbf {\bibinfo {volume}
  {436}},\ \bibinfo {pages} {333} (\bibinfo {year} {1947})}\BibitemShut
  {NoStop}%
\bibitem [{\citenamefont {Bliokh}\ and\ \citenamefont
  {Aiello}(2013)}]{Bliokh2013}%
  \BibitemOpen
  \bibfield  {author} {\bibinfo {author} {\bibfnamefont {K.~Y.}\ \bibnamefont
  {Bliokh}}\ and\ \bibinfo {author} {\bibfnamefont {A.}~\bibnamefont
  {Aiello}},\ }\href {\doibase 10.1088/2040-8978/15/1/014001} {\bibfield
  {journal} {\bibinfo  {journal} {{J}. {O}pt.}\ }\textbf {\bibinfo {volume}
  {15}},\ \bibinfo {pages} {014001} (\bibinfo {year} {2013})}\BibitemShut
  {NoStop}%
\bibitem [{\citenamefont {Keilmann}\ and\ \citenamefont
  {Hillenbrand}(2004)}]{Keilmann2004nfm}%
  \BibitemOpen
  \bibfield  {author} {\bibinfo {author} {\bibfnamefont {F.}~\bibnamefont
  {Keilmann}}\ and\ \bibinfo {author} {\bibfnamefont {R.}~\bibnamefont
  {Hillenbrand}},\ }\href {\doibase 10.1098/rsta.2003.1347} {\bibfield
  {journal} {\bibinfo  {journal} {{P}hil. {T}rans. {R}oy. {S}oc. {L}ondon,
  {S}er. {A}}\ }\textbf {\bibinfo {volume} {362}},\ \bibinfo {pages} {787}
  (\bibinfo {year} {2004})}\BibitemShut {NoStop}%
\bibitem [{\citenamefont {Atkin}\ \emph {et~al.}(2012)\citenamefont {Atkin},
  \citenamefont {Berweger}, \citenamefont {Jones},\ and\ \citenamefont
  {Raschke}}]{Atkin2012noi}%
  \BibitemOpen
  \bibfield  {author} {\bibinfo {author} {\bibfnamefont {J.~M.}\ \bibnamefont
  {Atkin}}, \bibinfo {author} {\bibfnamefont {S.}~\bibnamefont {Berweger}},
  \bibinfo {author} {\bibfnamefont {A.~C.}\ \bibnamefont {Jones}}, \ and\
  \bibinfo {author} {\bibfnamefont {M.~B.}\ \bibnamefont {Raschke}},\ }\href
  {\doibase 10.1080/00018732.2012.737982} {\bibfield  {journal} {\bibinfo
  {journal} {{A}dv. {P}hys.}\ }\textbf {\bibinfo {volume} {61}},\ \bibinfo
  {pages} {745} (\bibinfo {year} {2012})}\BibitemShut {NoStop}%
\bibitem [{\citenamefont {Wunsch}\ \emph {et~al.}(2006)\citenamefont {Wunsch},
  \citenamefont {Stauber}, \citenamefont {Sols},\ and\ \citenamefont
  {Guinea}}]{Wunsch2006}%
  \BibitemOpen
  \bibfield  {author} {\bibinfo {author} {\bibfnamefont {B.}~\bibnamefont
  {Wunsch}}, \bibinfo {author} {\bibfnamefont {T.}~\bibnamefont {Stauber}},
  \bibinfo {author} {\bibfnamefont {F.}~\bibnamefont {Sols}}, \ and\ \bibinfo
  {author} {\bibfnamefont {F.}~\bibnamefont {Guinea}},\ }\href {\doibase
  10.1088/1367-2630/8/12/318} {\bibfield  {journal} {\bibinfo  {journal} {{N}ew
  {J}. {P}hys.}\ }\textbf {\bibinfo {volume} {8}},\ \bibinfo {pages} {318}
  (\bibinfo {year} {2006})}\BibitemShut {NoStop}%
\bibitem [{\citenamefont {Hwang}\ and\ \citenamefont
  {Das~Sarma}(2007)}]{Hwang2007}%
  \BibitemOpen
  \bibfield  {author} {\bibinfo {author} {\bibfnamefont {E.~H.}\ \bibnamefont
  {Hwang}}\ and\ \bibinfo {author} {\bibfnamefont {S.}~\bibnamefont
  {Das~Sarma}},\ }\href {\doibase 10.1103/PhysRevB.75.205418} {\bibfield
  {journal} {\bibinfo  {journal} {{P}hys. {R}ev. {B}}\ }\textbf {\bibinfo
  {volume} {75}},\ \bibinfo {pages} {205418} (\bibinfo {year}
  {2007})}\BibitemShut {NoStop}%
\bibitem [{\citenamefont {LeBlanc}\ and\ \citenamefont
  {Carbotte}(2014)}]{LeBlanc2014dss}%
  \BibitemOpen
  \bibfield  {author} {\bibinfo {author} {\bibfnamefont {J.~P.~F.}\
  \bibnamefont {LeBlanc}}\ and\ \bibinfo {author} {\bibfnamefont {J.~P.}\
  \bibnamefont {Carbotte}},\ }\href {\doibase 10.1103/PhysRevB.89.035419}
  {\bibfield  {journal} {\bibinfo  {journal} {{P}hys. {R}ev. {B}}\ }\textbf
  {\bibinfo {volume} {89}},\ \bibinfo {pages} {035419} (\bibinfo {year}
  {2014})}\BibitemShut {NoStop}%
\bibitem [{\citenamefont {Stinson}\ \emph {et~al.}(2014)\citenamefont
  {Stinson}, \citenamefont {Wu}, \citenamefont {Jiang}, \citenamefont {Fei},
  \citenamefont {Rodin}, \citenamefont {Chapler}, \citenamefont {McLeod},
  \citenamefont {Castro~Neto}, \citenamefont {Lee}, \citenamefont {Fogler},\
  and\ \citenamefont {Basov}}]{Stinson2014}%
  \BibitemOpen
  \bibfield  {author} {\bibinfo {author} {\bibfnamefont {H.~T.}\ \bibnamefont
  {Stinson}}, \bibinfo {author} {\bibfnamefont {J.~S.}\ \bibnamefont {Wu}},
  \bibinfo {author} {\bibfnamefont {B.~Y.}\ \bibnamefont {Jiang}}, \bibinfo
  {author} {\bibfnamefont {Z.}~\bibnamefont {Fei}}, \bibinfo {author}
  {\bibfnamefont {A.~S.}\ \bibnamefont {Rodin}}, \bibinfo {author}
  {\bibfnamefont {B.~C.}\ \bibnamefont {Chapler}}, \bibinfo {author}
  {\bibfnamefont {A.~S.}\ \bibnamefont {McLeod}}, \bibinfo {author}
  {\bibfnamefont {A.}~\bibnamefont {Castro~Neto}}, \bibinfo {author}
  {\bibfnamefont {Y.~S.}\ \bibnamefont {Lee}}, \bibinfo {author} {\bibfnamefont
  {M.~M.}\ \bibnamefont {Fogler}}, \ and\ \bibinfo {author} {\bibfnamefont
  {D.~N.}\ \bibnamefont {Basov}},\ }\href {\doibase 10.1103/PhysRevB.90.014502}
  {\bibfield  {journal} {\bibinfo  {journal} {{P}hys. {R}ev. {B}}\ }\textbf
  {\bibinfo {volume} {90}},\ \bibinfo {pages} {014502} (\bibinfo {year}
  {2014})}\BibitemShut {NoStop}%
\bibitem [{\citenamefont {Sun}\ \emph {et~al.}(2015)\citenamefont {Sun},
  \citenamefont {Guti\'errez-Rubio}, \citenamefont {Basov},\ and\ \citenamefont
  {Fogler}}]{Sun2015hoh}%
  \BibitemOpen
  \bibfield  {author} {\bibinfo {author} {\bibfnamefont {Z.}~\bibnamefont
  {Sun}}, \bibinfo {author} {\bibfnamefont {A.}~\bibnamefont
  {Guti\'errez-Rubio}}, \bibinfo {author} {\bibfnamefont {D.~N.}\ \bibnamefont
  {Basov}}, \ and\ \bibinfo {author} {\bibfnamefont {M.~M.}\ \bibnamefont
  {Fogler}},\ }\href {\doibase 10.1021/acs.nanolett.5b00814} {\bibfield
  {journal} {\bibinfo  {journal} {{N}ano {L}ett.}\ }\textbf {\bibinfo {volume}
  {15}},\ \bibinfo {pages} {4455} (\bibinfo {year} {2015})}\BibitemShut
  {NoStop}%
\bibitem [{\citenamefont {Huerkamp}\ \emph {et~al.}(2011)\citenamefont
  {Huerkamp}, \citenamefont {Leskova}, \citenamefont {Maradudin},\ and\
  \citenamefont {Baumeier}}]{Huerkamp2011}%
  \BibitemOpen
  \bibfield  {author} {\bibinfo {author} {\bibfnamefont {F.}~\bibnamefont
  {Huerkamp}}, \bibinfo {author} {\bibfnamefont {T.~A.}\ \bibnamefont
  {Leskova}}, \bibinfo {author} {\bibfnamefont {A.~A.}\ \bibnamefont
  {Maradudin}}, \ and\ \bibinfo {author} {\bibfnamefont {B.}~\bibnamefont
  {Baumeier}},\ }\href {\doibase 10.1364/OE.19.015483} {\bibfield  {journal}
  {\bibinfo  {journal} {{O}pt. {E}xpr.}\ }\textbf {\bibinfo {volume} {19}},\
  \bibinfo {pages} {15483} (\bibinfo {year} {2011})}\BibitemShut {NoStop}%
\bibitem [{\citenamefont {Artmann}(1948)}]{Artmann1948}%
  \BibitemOpen
  \bibfield  {author} {\bibinfo {author} {\bibfnamefont {K.}~\bibnamefont
  {Artmann}},\ }\href {\doibase 10.1002/andp.19484370108} {\bibfield  {journal}
  {\bibinfo  {journal} {{A}nn. {P}hys.}\ }\textbf {\bibinfo {volume} {437}},\
  \bibinfo {pages} {87} (\bibinfo {year} {1948})}\BibitemShut {NoStop}%
\bibitem [{\citenamefont {Tamir}\ and\ \citenamefont
  {Oliner}(1963)}]{Tamir1963sew}%
  \BibitemOpen
  \bibfield  {author} {\bibinfo {author} {\bibfnamefont {T.}~\bibnamefont
  {Tamir}}\ and\ \bibinfo {author} {\bibfnamefont {A.}~\bibnamefont {Oliner}},\
  }\href {\doibase 10.1109/PROC.1963.1758} {\bibfield  {journal} {\bibinfo
  {journal} {{P}roc. {IEEE}}\ }\textbf {\bibinfo {volume} {51}},\ \bibinfo
  {pages} {317} (\bibinfo {year} {1963})}\BibitemShut {NoStop}%
\bibitem [{\citenamefont {Tamir}\ and\ \citenamefont
  {Bertoni}(1971)}]{Tamir1971ldo}%
  \BibitemOpen
  \bibfield  {author} {\bibinfo {author} {\bibfnamefont {T.}~\bibnamefont
  {Tamir}}\ and\ \bibinfo {author} {\bibfnamefont {H.~L.}\ \bibnamefont
  {Bertoni}},\ }\href {\doibase 10.1364/JOSA.61.001397} {\bibfield  {journal}
  {\bibinfo  {journal} {{J}. {O}pt. {S}oc. {A}m.}\ }\textbf {\bibinfo {volume}
  {61}},\ \bibinfo {pages} {1397} (\bibinfo {year} {1971})}\BibitemShut
  {NoStop}%
\bibitem [{\citenamefont {Chuang}(1986)}]{Chuang1986lso}%
  \BibitemOpen
  \bibfield  {author} {\bibinfo {author} {\bibfnamefont {S.~L.}\ \bibnamefont
  {Chuang}},\ }\href {\doibase 10.1364/JOSAA.3.000593} {\bibfield  {journal}
  {\bibinfo  {journal} {{J}. {O}pt. {S}oc. {A}m. {A}}\ }\textbf {\bibinfo
  {volume} {3}},\ \bibinfo {pages} {593} (\bibinfo {year} {1986})}\BibitemShut
  {NoStop}%
\bibitem [{\citenamefont {Yin}\ \emph {et~al.}(2004)\citenamefont {Yin},
  \citenamefont {Hesselink}, \citenamefont {Liu}, \citenamefont {Fang},\ and\
  \citenamefont {Zhang}}]{yin2004}%
  \BibitemOpen
  \bibfield  {author} {\bibinfo {author} {\bibfnamefont {X.}~\bibnamefont
  {Yin}}, \bibinfo {author} {\bibfnamefont {L.}~\bibnamefont {Hesselink}},
  \bibinfo {author} {\bibfnamefont {Z.}~\bibnamefont {Liu}}, \bibinfo {author}
  {\bibfnamefont {N.}~\bibnamefont {Fang}}, \ and\ \bibinfo {author}
  {\bibfnamefont {X.}~\bibnamefont {Zhang}},\ }\href {\doibase
  http://dx.doi.org/10.1063/1.1775294} {\bibfield  {journal} {\bibinfo
  {journal} {{A}ppl. {P}hys. {L}ett.}\ }\textbf {\bibinfo {volume} {85}},\
  \bibinfo {pages} {372} (\bibinfo {year} {2004})}\BibitemShut {NoStop}%
\bibitem [{\citenamefont {Feurer}\ \emph {et~al.}(2007)\citenamefont {Feurer},
  \citenamefont {Stoyanov}, \citenamefont {Ward}, \citenamefont {Vaughan},
  \citenamefont {Statz},\ and\ \citenamefont {Nelson}}]{Feurer2007tp}%
  \BibitemOpen
  \bibfield  {author} {\bibinfo {author} {\bibfnamefont {T.}~\bibnamefont
  {Feurer}}, \bibinfo {author} {\bibfnamefont {N.~S.}\ \bibnamefont
  {Stoyanov}}, \bibinfo {author} {\bibfnamefont {D.~W.}\ \bibnamefont {Ward}},
  \bibinfo {author} {\bibfnamefont {J.~C.}\ \bibnamefont {Vaughan}}, \bibinfo
  {author} {\bibfnamefont {E.~R.}\ \bibnamefont {Statz}}, \ and\ \bibinfo
  {author} {\bibfnamefont {K.~A.}\ \bibnamefont {Nelson}},\ }\href {\doibase
  10.1146/annurev.matsci.37.052506.084327} {\bibfield  {journal} {\bibinfo
  {journal} {{A}nn. {R}ev. {M}ater. {R}es.}\ }\textbf {\bibinfo {volume}
  {37}},\ \bibinfo {pages} {317} (\bibinfo {year} {2007})}\BibitemShut
  {NoStop}%
\bibitem [{\citenamefont {Jin}\ \emph {et~al.}(2013)\citenamefont {Jin},
  \citenamefont {Kumar}, \citenamefont {Hung~Fung}, \citenamefont {Xu},\ and\
  \citenamefont {Fang}}]{Jin2013tpf}%
  \BibitemOpen
  \bibfield  {author} {\bibinfo {author} {\bibfnamefont {D.}~\bibnamefont
  {Jin}}, \bibinfo {author} {\bibfnamefont {A.}~\bibnamefont {Kumar}}, \bibinfo
  {author} {\bibfnamefont {K.}~\bibnamefont {Hung~Fung}}, \bibinfo {author}
  {\bibfnamefont {J.}~\bibnamefont {Xu}}, \ and\ \bibinfo {author}
  {\bibfnamefont {N.~X.}\ \bibnamefont {Fang}},\ }\href {\doibase
  10.1063/1.4807762} {\bibfield  {journal} {\bibinfo  {journal} {{A}ppl.
  {P}hys. {L}ett.}\ }\textbf {\bibinfo {volume} {102}},\ \bibinfo {pages}
  {201118} (\bibinfo {year} {2013})}\BibitemShut {NoStop}%
\bibitem [{\citenamefont {Cha}\ \emph {et~al.}(2013)\citenamefont {Cha},
  \citenamefont {Koski}, \citenamefont {Huang}, \citenamefont {Wang},
  \citenamefont {Luo}, \citenamefont {Kong}, \citenamefont {Yu}, \citenamefont
  {Fan}, \citenamefont {Brongersma},\ and\ \citenamefont {Cui}}]{Cha2013}%
  \BibitemOpen
  \bibfield  {author} {\bibinfo {author} {\bibfnamefont {J.~J.}\ \bibnamefont
  {Cha}}, \bibinfo {author} {\bibfnamefont {K.~J.}\ \bibnamefont {Koski}},
  \bibinfo {author} {\bibfnamefont {K.~C.~Y.}\ \bibnamefont {Huang}}, \bibinfo
  {author} {\bibfnamefont {K.~X.}\ \bibnamefont {Wang}}, \bibinfo {author}
  {\bibfnamefont {W.}~\bibnamefont {Luo}}, \bibinfo {author} {\bibfnamefont
  {D.}~\bibnamefont {Kong}}, \bibinfo {author} {\bibfnamefont {Z.}~\bibnamefont
  {Yu}}, \bibinfo {author} {\bibfnamefont {S.}~\bibnamefont {Fan}}, \bibinfo
  {author} {\bibfnamefont {M.~L.}\ \bibnamefont {Brongersma}}, \ and\ \bibinfo
  {author} {\bibfnamefont {Y.}~\bibnamefont {Cui}},\ }\href {\doibase
  10.1021/nl402937g} {\bibfield  {journal} {\bibinfo  {journal} {{N}ano
  {L}ett.}\ }\textbf {\bibinfo {volume} {13}},\ \bibinfo {pages} {5913}
  (\bibinfo {year} {2013})}\BibitemShut {NoStop}%
\bibitem [{\citenamefont {Ou}\ \emph {et~al.}(2014)\citenamefont {Ou},
  \citenamefont {So}, \citenamefont {Adamo}, \citenamefont {Sulaev},
  \citenamefont {Wang},\ and\ \citenamefont {Zheludev}}]{Ou2014}%
  \BibitemOpen
  \bibfield  {author} {\bibinfo {author} {\bibfnamefont {J.-Y.}\ \bibnamefont
  {Ou}}, \bibinfo {author} {\bibfnamefont {J.-K.}\ \bibnamefont {So}}, \bibinfo
  {author} {\bibfnamefont {G.}~\bibnamefont {Adamo}}, \bibinfo {author}
  {\bibfnamefont {A.}~\bibnamefont {Sulaev}}, \bibinfo {author} {\bibfnamefont
  {L.}~\bibnamefont {Wang}}, \ and\ \bibinfo {author} {\bibfnamefont {N.~I.}\
  \bibnamefont {Zheludev}},\ }\href {\doibase 10.1038/ncomms6139} {\bibfield
  {journal} {\bibinfo  {journal} {{N}at. {C}omm.}\ }\textbf {\bibinfo {volume}
  {5}},\ \bibinfo {pages} {5139} (\bibinfo {year} {2014})}\BibitemShut
  {NoStop}%
\bibitem [{\citenamefont {Jacob}\ \emph {et~al.}(2006)\citenamefont {Jacob},
  \citenamefont {Alekseyev},\ and\ \citenamefont {Narimanov}}]{Jacob2006}%
  \BibitemOpen
  \bibfield  {author} {\bibinfo {author} {\bibfnamefont {Z.}~\bibnamefont
  {Jacob}}, \bibinfo {author} {\bibfnamefont {L.~V.}\ \bibnamefont
  {Alekseyev}}, \ and\ \bibinfo {author} {\bibfnamefont {E.}~\bibnamefont
  {Narimanov}},\ }\href {\doibase 10.1364/OE.14.008247} {\bibfield  {journal}
  {\bibinfo  {journal} {{O}pt. {E}xpress}\ }\textbf {\bibinfo {volume} {14}},\
  \bibinfo {pages} {8247} (\bibinfo {year} {2006})}\BibitemShut {NoStop}%
\bibitem [{\citenamefont {Salandrino}\ and\ \citenamefont
  {Engheta}(2006)}]{Salandrino2006ffs}%
  \BibitemOpen
  \bibfield  {author} {\bibinfo {author} {\bibfnamefont {A.}~\bibnamefont
  {Salandrino}}\ and\ \bibinfo {author} {\bibfnamefont {N.}~\bibnamefont
  {Engheta}},\ }\href {\doibase 10.1103/PhysRevB.74.075103} {\bibfield
  {journal} {\bibinfo  {journal} {{P}hys. {R}ev. {B}}\ }\textbf {\bibinfo
  {volume} {74}},\ \bibinfo {pages} {075103} (\bibinfo {year}
  {2006})}\BibitemShut {NoStop}%
\bibitem [{\citenamefont {Liu}\ \emph {et~al.}(2007)\citenamefont {Liu},
  \citenamefont {Lee}, \citenamefont {Xiong}, \citenamefont {Sun},\ and\
  \citenamefont {Zhang}}]{Liu2007ffo}%
  \BibitemOpen
  \bibfield  {author} {\bibinfo {author} {\bibfnamefont {Z.}~\bibnamefont
  {Liu}}, \bibinfo {author} {\bibfnamefont {H.}~\bibnamefont {Lee}}, \bibinfo
  {author} {\bibfnamefont {Y.}~\bibnamefont {Xiong}}, \bibinfo {author}
  {\bibfnamefont {C.}~\bibnamefont {Sun}}, \ and\ \bibinfo {author}
  {\bibfnamefont {X.}~\bibnamefont {Zhang}},\ }\href {\doibase
  10.1126/science.1137368} {\bibfield  {journal} {\bibinfo  {journal}
  {{S}cience}\ }\textbf {\bibinfo {volume} {315}},\ \bibinfo {pages} {1686}
  (\bibinfo {year} {2007})}\BibitemShut {NoStop}%
\bibitem [{\citenamefont {Zhang}\ \emph {et~al.}(2012)\citenamefont {Zhang},
  \citenamefont {Andreev}, \citenamefont {Fei}, \citenamefont {McLeod},
  \citenamefont {Dominguez}, \citenamefont {Thiemens}, \citenamefont
  {Castro-Neto}, \citenamefont {Basov},\ and\ \citenamefont
  {Fogler}}]{Zhang2012}%
  \BibitemOpen
  \bibfield  {author} {\bibinfo {author} {\bibfnamefont {L.~M.}\ \bibnamefont
  {Zhang}}, \bibinfo {author} {\bibfnamefont {G.~O.}\ \bibnamefont {Andreev}},
  \bibinfo {author} {\bibfnamefont {Z.}~\bibnamefont {Fei}}, \bibinfo {author}
  {\bibfnamefont {A.~S.}\ \bibnamefont {McLeod}}, \bibinfo {author}
  {\bibfnamefont {G.}~\bibnamefont {Dominguez}}, \bibinfo {author}
  {\bibfnamefont {M.}~\bibnamefont {Thiemens}}, \bibinfo {author}
  {\bibfnamefont {A.~H.}\ \bibnamefont {Castro-Neto}}, \bibinfo {author}
  {\bibfnamefont {D.~N.}\ \bibnamefont {Basov}}, \ and\ \bibinfo {author}
  {\bibfnamefont {M.~M.}\ \bibnamefont {Fogler}},\ }\href {\doibase
  10.1103/PhysRevB.85.075419} {\bibfield  {journal} {\bibinfo  {journal}
  {{P}hys. {R}ev. {B}}\ }\textbf {\bibinfo {volume} {85}},\ \bibinfo {pages}
  {075419} (\bibinfo {year} {2012})}\BibitemShut {NoStop}%
\bibitem [{\citenamefont {Jiang}\ \emph {et~al.}()\citenamefont {Jiang},
  \citenamefont {Zhang}, \citenamefont {Castro~Neto}, \citenamefont {Basov},\
  and\ \citenamefont {Fogler}}]{Jiang2015gsm}%
  \BibitemOpen
  \bibfield  {author} {\bibinfo {author} {\bibfnamefont {B.-Y.}\ \bibnamefont
  {Jiang}}, \bibinfo {author} {\bibfnamefont {L.~M.}\ \bibnamefont {Zhang}},
  \bibinfo {author} {\bibfnamefont {A.~H.}\ \bibnamefont {Castro~Neto}},
  \bibinfo {author} {\bibfnamefont {D.~N.}\ \bibnamefont {Basov}}, \ and\
  \bibinfo {author} {\bibfnamefont {M.~M.}\ \bibnamefont {Fogler}},\ }\href
  {http://arxiv.org/abs/1503.00221} {\enquote {\bibinfo {title} {{G}eneralized
  spectral method for near-field optical microscopy},}\ }\bibinfo {note}
  {{a}r{X}iv:1503.00221}\BibitemShut {NoStop}%
\bibitem [{\citenamefont {McLeod}\ \emph {et~al.}(2014)\citenamefont {McLeod},
  \citenamefont {Kelly}, \citenamefont {Goldflam}, \citenamefont {Gainsforth},
  \citenamefont {Westphal}, \citenamefont {Dominguez}, \citenamefont
  {Thiemens}, \citenamefont {Fogler},\ and\ \citenamefont
  {Basov}}]{McLeod2014mqt}%
  \BibitemOpen
  \bibfield  {author} {\bibinfo {author} {\bibfnamefont {A.~S.}\ \bibnamefont
  {McLeod}}, \bibinfo {author} {\bibfnamefont {P.}~\bibnamefont {Kelly}},
  \bibinfo {author} {\bibfnamefont {M.~D.}\ \bibnamefont {Goldflam}}, \bibinfo
  {author} {\bibfnamefont {Z.}~\bibnamefont {Gainsforth}}, \bibinfo {author}
  {\bibfnamefont {A.~J.}\ \bibnamefont {Westphal}}, \bibinfo {author}
  {\bibfnamefont {G.}~\bibnamefont {Dominguez}}, \bibinfo {author}
  {\bibfnamefont {M.~H.}\ \bibnamefont {Thiemens}}, \bibinfo {author}
  {\bibfnamefont {M.~M.}\ \bibnamefont {Fogler}}, \ and\ \bibinfo {author}
  {\bibfnamefont {D.~N.}\ \bibnamefont {Basov}},\ }\href {\doibase
  10.1103/PhysRevB.90.085136} {\bibfield  {journal} {\bibinfo  {journal}
  {{P}hys. {R}ev. {B}}\ }\textbf {\bibinfo {volume} {90}},\ \bibinfo {pages}
  {085136} (\bibinfo {year} {2014})}\BibitemShut {NoStop}%
\end{thebibliography}%
\end{document}